\documentclass[12pt]{article}
\usepackage{graphicx}

\begin{document}

\begin{center}
{\large {\bf Transverse Wave Propagation in Relativistic Two-fluid Plasmas around Reissner-Nordstr\"{o}m-de Sitter Black Hole}}\\
\vspace{1.5cm}
M. Atiqur Rahman\footnote{E-mail: $atirubd@yahoo.com$}and
M. Hossain Ali
\footnote{E-mail: $m_-hossain_-ali_-bd@yahoo.com$ (Corresponding author)}\\
{\it Department of Applied Mathematics,\\ Rajshahi University,\\ Rajshahi - 6205, Bangladesh}
\end{center}
\vspace{1.5cm}
\centerline{\bf Abstract}
The transverse electromagnetic waves propagating in a relativistic two-fluid plasma influenced by the gravitational field of the Reissner-Nordstr\"{o}m-de Sitter black hole has been investigated exploiting 3+1 split of spacetime. Reformulating the two-fluid equations, the set of simultaneous linear equations for the perturbations have been derived. Using a local approximation, the one-dimensional radial propagation of Alfv\'{e}n and high frequency electromagnetic waves are investigated. The dispersion relation for these waves is obtained and solved numerically for the wave number.\\
\vspace{0.1cm}\\
\textbf{Keywords}: Two-fluid plasma, Alfv\'en and high frequency electromagnetic waves, Charged black hole in de Sitter space.\\
\textbf{PACS number(s)}: 95.30.Sf, 95.30.Qd, 97.60.Lf
\vspace{0.2cm}\\

\newpage

\section{Introduction}\label{sec1}
Black holes belong to the most fascinating objects predicted by Einstein's theory of gravitation and are still mysterious \cite{one}. Physicists are grappling the theory of black holes, while astronomers are searching for real-life examples of black holes in the universe \cite{two}. However, the theory of general relativity and its application to the plasma close to the black hole horizon have remained esoteric, and little concrete astrophysical impact has been felt. Within $3R_S$ (3 Schwarzschild radii) it is possible to have plasma \cite{three,four,five,six}. The plasma in the black hole environment may act as a fluid and black holes greatly affect the surrounding plasma medium (which is highly magnetized) with their enormous gravitational fields. Hence plasma physics in the vicinity of a black hole has become a subject of great interest in astrophysics. In the immediate neighborhood of a black hole general relativity applies. It is therefore of interest to investigate plasma physics problems in the context of general relativity.

In recent years, considerable attention has been concentrated on the study of black holes in de Sitter (dS) spaces. The motivation behind it is based on two aspects: first, several different types of astrophysical observations indicate that our universe is in a phase of accelerating expansion \cite{seven,eight,nine,ten,eleven,twelve,thirteen,fourteen,fifteen,sixteen,seventeen,eighteen,nineteen,twenty,twenty one,twenty two,twenty three,twenty four,twenty five,twenty six,twenty seven,twenty eight,twenty nine}, associated with which is a positive cosmological constant. Our universe therefore might approach a dS phase in the far future \cite{thirty,thirty one,thirty two,thirty three,thirty four,thirty five,thirty six}. Secondly, like the AdS/CFT correspondence, an interesting proposal, the so-called dS/CFT correspondence, has been suggested that there is a dual relation between quantum gravity on a dS space and Euclidean conformal field theory (CFT) on a boundary of dS space \cite{thirty seven,thirty eight,thirty nine,fourty,fourty one,fourty two}. In view of these reasons, the study of transverse wave propagation in relativistic two-fluid plasma in the vicinity of a black hole in dS space is interesting.

A covariant formulation of the theory based on the fluid equations of general relativity and Maxwell's equations in curved spacetime have so far proved unproductive because of the curvature of four-dimensional spacetime in the region surrounding a black hole. Thorne, Price, and Macdonald (TPM) \cite{fourty three,fourty four,fourty five,fourty six} developed a method of 3+1 formulation of general relativity in which the event horizon of black holes was replaced with a membrane endowed with electric charge, electrical conductivity, and finite temperature and entropy so that the physics outside the event horizon turns out to be very much simpler than it would be using the standard covariant approach of general relativity. Buzzi et al. \cite{fourty seven,fourty eight} employed the TPM formulation to develope a general relativistic version of two-fluid plasma and investigated the nature of plasma in the vicinity of Schwarzschild black hole. Exploiting the formalism of Buzzi et al., we investigate the transverse wave propagation in the two-fluid plasma around the Reissner-Nordstr\"{o}m-de Sitter (RNdS) black hole which is the Schwarzschild black hole generalized with a charge parameter and a positive cosmological constant, described by the metric
\begin{eqnarray}
ds^2=-f(r)dt^2+\frac{dr^2}{f(r)}+r^2(d\theta ^2+{\rm sin}^2\theta d\varphi ^2),\label{eq1}
\end{eqnarray}
where the metric function $f(r)$ is given by
\begin{eqnarray}
f(r)=1-\frac{2M}{r}+\frac{q^2}{r^2}-\frac {r^2}{\ell^2},\label{eq2}
\end{eqnarray}
$M$ being the mass, and $q$ the total charge (electric plus magnetic) in the background of the static de Sitter space. The positive cosmological constant $\Lambda $ is usually written as $\Lambda =3/\ell^2$ with $\ell$ the cosmological radius. The spacetime causal structure depends strongly on the singularities of the metric given by the zeros of $f(r)$. Depending on the parameters $M$, $q$, and $\Lambda $, the function $f(r)$ may have three, two, or even no real positive zeros. For the RNdS black hole case we are interested in which $f(r)$ has three real, positive roots ($r_-<r_+<r_c$), and a real negative root $r_n=-(r_c+r_++r_-)$. The horizons at $r_-$, $r_+$, and $r_c$, are called the inner (Cauchy), outer (event), and cosmological horizons, respectively. The black hole parameters $M$, $q$, and $\ell$ are related to the roots by
\begin{eqnarray}
{r^2_+}+{r^2_c}+{r_+}{r_c}-{r_-}{r_n}&=&\ell^2,\nonumber\\
(r_++r_c)(r_+r_c-r_-r_n)&=&2\ell^2M,\nonumber\\
{r_+}{r_c}{r_-}{r_n}&=&-\ell^2 q^2.\label{eq3}
\end{eqnarray}
For the zeros of $f(r)$, we have
\begin{equation}
r^4-\ell^2r^2+2M\ell^2r-\ell^2q^2=0. \label{eq4}
\end{equation}
Positions of the black hole horizons are given by
\begin{equation}
r_\pm =\frac{\ell}{\sqrt{3}}\sin\left[\frac{1}{3}\sin^{-1} \frac{3\sqrt{3}M}{\ell\sqrt{1+\frac{4q^2}{\ell^2}}}\right]\left(1\pm \sqrt{1-\frac{\sqrt{3}q^2}{\ell M}\frac{2}{1+\zeta}\csc\left[\frac{1}{3}\sin^{-1} \frac{3\sqrt{3}M}{\ell\sqrt{1+\frac{4q^2} {\ell^2}}}\right]}\right)\label{eq5},
\end{equation}
where
\begin{equation}
\zeta=\sqrt{1-\frac{4q^2}{3M^2}\sin^2 \left[\frac{1}{3}\sin^{-1}\frac{3\sqrt{3}M} {\ell\sqrt{1+\frac{4q^2}{\ell^2}}}\right]}\label{eq6}.
\end{equation}
The cosmological horizon is located at
\begin{equation}
r_c=\frac{\ell}{\sqrt{3}}\sin\left[\frac{1}{3}\sin^{-1}\frac{ 3\sqrt{3}M}{\ell\sqrt{1+\frac{4q^2}{\ell^2}}}\right]
\left(\sqrt{1+\frac{1+\zeta}{2}\frac{3\sqrt{3}M}{ \ell}\csc^3\left[\frac{1}{3}\sin^{-1}\frac{3\sqrt{3}M}{ \ell\sqrt{1+\frac{4q^2}{\ell^2}}}\right]}-1\right)\label{eq7}.
\end{equation}
Since $27\frac {M^2}{\ell^2}<1$, therefore ${3\sqrt{3}\frac {M}{\ell}}/{\sqrt{1+\frac{4q^2}{\ell^2}}}<1$, and we obtain
\begin{equation}
r_\pm=\frac{M}{\xi}\left(1\pm\sqrt{1-\frac{q^2}{ M^2}\frac{2\xi}{1+\zeta}}\right)\label{eq8},
\end{equation}
where
\begin{equation}
\xi=\frac{1}{\eta}\sqrt{1+\frac{4q^2}{\ell^2}}, \qquad\eta=1+\frac{4M^2}{\ell^2(1+\frac{4q^2}{\ell^2})}+ \cdots\label{eq9}.
\end{equation}
It gives the the RN black hole horizons for $\ell\rightarrow\infty$ and the Schwarzschild-de Sitter black hole horizon for $q=0$. The metric (\ref{eq1}) represents an interesting asymptotically de Sitter extreme RN black hole (or, \lq \lq cold\rq \rq black hole) for $q^2=\frac{2\xi}{1+\zeta}M^2$, while for $q^2>\frac{2\xi}{1+\zeta}M^2$ it does not represent any black hole but an unphysical naked singularity.

The black hole horizon $r_+$ and cosmological horizon $r_c$ are not in thermal equilibrium because the time periods in the Euclidean section required to avoid a conical singularity at both do not match in general. However, there are two families of RNdS solutions with the same black hole and cosmological horizon temperatures, representing the \lq \lq lukewarm\rq \rq black hole and charged Nariai black hole \cite{fourty nine,fifty}. The lukewarm black holes are characterized by $q^2=M^2$, which is the condition of stable endpoints of the evaporation process. Its three horizons do not coincide and cosmological horizon and black hole horizon satisfy \cite{fifty one,fifty two}
\begin{equation}
M^2=q^2=\frac{r_+^2r_c^2}{(r_++r_c)^2},\qquad \ell=r_++r_c\label{eq10}.
\end{equation}
The black hole is colder than the lukewarm solution for $q^2>M^2$ and it will then absorb radiation from the cosmological horizon. As $q^2$ increases relative to $M^2$ the inner and outer horizons eventually coincide. The resulting hole then becomes extremal (or \lq \lq cold\rq \rq) RNdS black hole with the mass, charge, horizon radius and cosmological constant satisfying the following relations \cite{fifty two}:
\begin{equation}
\frac{M}{r_+(r_c+r_+)^2}=\frac{q^2}{r_cr_+^2(r_c+2r_+)} =\frac{1}{\ell^2}=\frac{1}{r_c^2+2r_+r_c+3r_+^2}\label{eq11}.
\end{equation}

As $M$ increases relative to $|q|$ with $M>0$, the outer black hole and cosmological horizons come closer together. When these horizons coincide at $r_h$, where $f(r_h)=f^\prime(r_h)=0$:
\begin{equation}
r_h=\frac{3}{2}M\left(1\pm\sqrt{1- \frac{8q^2}{9M^2}}\right)\label{eq12},
\end{equation}
it then becomes the charged Nariai solution, which is the largest charged asymptotically dS black hole. For a given $q$, it has the maximal mass $M_N$. The charged Nariai solution and lukewarm solution join together for the critical value $q^2=M_c^2=\ell^2/16$. If $M_{min}$ is the mass of extremal RNdS black hole, the range for mass parameter of a black hole in dS space is $M_{min}\leq M\leq M_N$, referred to as the \lq \lq undermassive\rq \rq case. The metric (\ref{eq1}) describes  a naked singularity, if this limit is exceeded.

If the three horizons coincide, i.e. $r_c=r_+=r_-$, the RNdS black hole becomes an \lq \lq ultracold\rq \rq black hole with horizon at $r_{ucd}$, where $f(r_{ucd})=f^\prime(r_{ucd})=f^{\prime\prime}(r_{ucd})=0$. The mass, charge and cosmological constant are related according to
\begin{equation}
M=\frac{2}{3}r_{ucd},\qquad q^2=\frac{1}{2}r^2_{ucd},\qquad \ell^2=6r^2_{ucd}\label{eq13}.
\end{equation}
This configuration simultaneously maximizes the values of $M$ and $q^2$ for any given positive value of the cosmological constant $\Lambda$.

The organization of this paper is as follows: In section \ref{sec2}, we summarize the 3+1 formulation of general relativity and Maxwell's equations in curved spacetime. In section \ref{sec3}, we describe the horizon governing equations of two-fluid plasma in the RNdS black hole spacetime. In section \ref{sec4}, we consider the wave propagation in the radial $z$ direction, and linearize the equations for wave propagation in section \ref{sec5} by giving a small perturbation to fields and fluid parameters. In section \ref{sec6}, we derive the dispersion relation for the transverse waves and discuss the local approximation used to obtain numerical solution of it. In section \ref{sec7}, we discuss our numerical solution modes and present the results in section \ref{sec8}. Finally, we give our concluding remarks in section \ref{sec9}. Throughout the paper we use natural units: $G=c=k_B=1$.

\section{3+1 Formulation in Spacetime}\label{sec2}

An absolute three-dimensional space defined by the hypersurfaces of constant universal time $t$ is described by the metric
\begin{equation}
ds^2=\textrm{g}_{ij}dx^idx^j=\frac{1}{f(r)}dr^2+r^2(d\theta ^2+{\rm sin}^2\theta d\varphi ^2)\label{eq14}.
\end{equation}
The indices $i, j$ range over $1,2,3$ and refer to coordinates in absolute space. The fiducial observers (FIDOs), that is, the observers at rest with respect to this absolute space, measure their proper time $\tau $ using clocks that they carry with them and make local measurements of physical quantities. Then all their measured quantities are defined as FIDO locally measured quantities and all rates measured by them are measured using FIDO proper time. The FIDOs use a local Cartesian coordinate system with unit basis vectors tangent to the coordinate lines:
\begin{equation}
{\bf e}_{\hat r}=\sqrt{f(r)}\frac{\partial }{\partial r},\hspace{1cm}{\bf e}_{\hat \theta }=\frac{1}{r}\frac{\partial }{\partial \theta },\hspace{1cm}{\bf e}_{\hat \varphi }=\frac{1}{r{\rm sin}\theta }\frac{\partial }{\partial \varphi }\label{eq15}.
\end{equation}
For a spacetime viewpoint rather than a 3+1 split of spacetime, the set of orthonormal vectors also includes the basis vector for the time coordinate given by
\begin{equation}
{\bf e}_{\hat 0}=\frac{d}{d\tau }=\frac{1}{\alpha }\frac{\partial }{\partial t}\label{eq16},
\end{equation}
where $\alpha $ is the lapse function (or redshift factor) defined by
\begin{equation}
\alpha (r)\equiv \frac{d\tau }{dt}=\left(1-\frac{2M}{r}+\frac{q^2}{r^2}-\frac{r^2}{ \ell^2}\right)^{\frac{1}{2}}\label{eq17}.
\end{equation}

The gravitational acceleration felt by a FIDO is given by \cite{fourty three,fourty four,fourty five,fourty six}
\begin{equation}
{\bf a}=-\nabla {\rm ln}\alpha =-\frac{1}{\alpha }\left({\frac{M}{r^2}}-\frac{q^2}{r^3}-{\frac{r }{\ell^2}}\right){\bf e}_{\hat r}\label{eq18},
\end{equation}
while the rate of change of any scalar physical quantity or any three-dimensional vector or tensor, as measured by a FIDO, is defined by the derivative
\begin{equation}
\frac{D}{D\tau }\equiv \left(\frac{1}{\alpha }\frac{\partial }{\partial t}+{\bf v}\cdot \nabla \right)\label{eq19},
\end{equation}
$\bf v$ being the velocity of a fluid as measured locally by a FIDO.

\section{Two-fluid Plasma Governing Equations}\label{sec3}
We consider two-component plasma such as an electron-positron or electron-ion. In the TPM notation, the continuity equation for each of the fluid species is given by
\begin{equation}
\frac{\partial }{\partial t}(\gamma _sn_s)+\nabla \cdot (\alpha \gamma _sn_s{\bf v}_s)=0\label{eq20},
\end{equation}
where $s$ is $1$ for electrons and $2$ for positrons (or ions). For a perfect relativistic fluid of species $s$ in three-dimensions, the energy density $\epsilon _s$, the momentum density ${\bf S}_s$, and stress-energy tensor $W_s^{jk}$ are, respectively, given by
\begin{eqnarray}
\epsilon _s=\gamma _s^2(\varepsilon _s+P_s{\bf v}_s^2),\hspace{0.6cm}{\bf S}_s=\gamma _s^2(\varepsilon _s+P_s){\bf v}_s,\hspace{0.6cm} W_s^{ij}=\gamma _s^2(\varepsilon _s+P_s)v_s^jv_s^k+P_s\textrm{g}^{jk}\label{eq21},
\end{eqnarray}
where ${\bf v}_s$ is the fluid velocity, $n_s$ is the number density, $P_s$ is the pressure, and $\varepsilon _s$ is the total energy density defined by
\begin{equation}
\varepsilon _s=m_sn_s+P_s/(\gamma _{\textrm{g}}-1)\label{eq22}.
\end{equation}
The gas constant $\gamma _\textrm{g}$ is $4/3$ for $T\rightarrow \infty $ and $5/3$ for $T\rightarrow 0$.

When the two-fluid plasma couples to the electromagnetic fields, Maxwell's equations take the following form:
\begin{eqnarray}
\nabla \cdot {\bf B}&=&0,\label{eq23}\\
\nabla \cdot {\bf E}&=&4\pi \sigma ,\label{eq24}\\
\frac{\partial {\bf B}}{\partial t}&=&-\nabla \times (\alpha {\bf E}),\label{eq25}\\
\frac{\partial {\bf E}}{\partial t}&=&\nabla \times (\alpha {\bf B})-4\pi \alpha {\bf J},\label{eq26}
\end{eqnarray}
where the charge and current densities are respectively defined by
\begin{equation}
\sigma =\sum_s\gamma _sq_sn_s,\quad {\bf J}=\sum_s\gamma _sq_sn_s{\bf v}_s\label{eq27}.
\end{equation}

The fluid quantities in (\ref{eq21}) take the following form in the electromagnetic field:
\begin{eqnarray}
\epsilon _s=\frac{1}{8\pi }({\bf E}^2+{\bf B}^2),\qquad {\bf S}_s=\frac{1}{4\pi }{\bf E}\times {\bf B},\nonumber\\
W_s^{jk}=\frac{1}{8\pi }({\bf E}^2+{\bf B}^2)\textrm{g}^{jk}-\frac{1}{4\pi }(E^jE^k+B^jB^k)\label{eq28}.
\end{eqnarray}
Energy and momentum conservation equations are respectively expressed by \cite{fourty three,fourty four,fourty five,fourty six}
\begin{eqnarray}
\frac{1}{\alpha }\frac{\partial }{\partial t}\epsilon _s=-\nabla \cdot {\bf S}_s+2{\bf a}\cdot {\bf S}_s,\label{eq29}\\
\frac{1}{\alpha }\frac{\partial }{\partial t}{\bf S}_s=\epsilon _s{\bf a}-\frac{1}{\alpha }\nabla \cdot (\alpha {\stackrel{\leftrightarrow }{\bf W}}_s)\label{eq30},
\end{eqnarray}
which, using (\ref{eq22})-(\ref{eq26}), can be written for each species $s$ in the form
\begin{eqnarray}
\frac{1}{\alpha }\frac{\partial }{\partial t}P_s-\frac{1}{\alpha }\frac{\partial }{\partial t}[\gamma _s^2(\varepsilon _s+P_s)]-\nabla \cdot [\gamma _s^2(\varepsilon _s+P_s){\bf v}_s]\nonumber\\
+\gamma _sq_sn_s{\bf E}\cdot {\bf v}_s+2\gamma _s^2(\varepsilon _s+P_s){\bf a}\cdot {\bf v}_s=0,\label{eq31}
\end{eqnarray}
\begin{eqnarray}
\gamma _s^2(\varepsilon _s+P_s)\left(\frac{1}{\alpha }\frac{\partial }{\partial t} +{\bf v}_s\cdot \nabla \right){\bf v}_s+\nabla P_s-\gamma _sq_sn_s({\bf E}+{\bf v}_s\times{\bf B})\nonumber\\
+{\bf v}_s\left(\gamma _sq_sn_s{\bf E}\cdot {\bf v}_s+\frac{1}{\alpha }\frac{\partial }{\partial t}P_s\right)+\gamma _s^2(\varepsilon _s+P_s)[{\bf v}_s({\bf v}_s\cdot {\bf a})-{\bf a}]=0\label{eq32}.
\end{eqnarray}
Although these equations are valid in a FIDO frame, they reduce to the corresponding special relativistic case for $\alpha =1$ \cite{fifty three}, which are valid in a frame in which both fluids are at rest. The transformation from the FIDO frame to the comoving (fluid) frame involves a boost velocity, which is simply the freefall velocity onto the black hole, given by
\begin{equation}
v_{\rm ff}=(1-\alpha ^2)^{\frac{1}{2}}\label{eq33}.
\end{equation}
Then the relativistic Lorentz factor $\gamma _{\rm boost}\equiv (1-v_{\rm ff}^2)^{-\frac{1}{2}}=1/\alpha $.

The ion temperature profile is closely adiabatic and it approaches $10^{12}\,\textrm{K}$ near the horizon \cite{fifty four}. Far from the (event) horizon electron (positron) temperatures are essentially equal to the ion temperatures, but closer to the horizon the electrons are progressively cooled to about $10^8-10^9\,\textrm{K}$ by mechanisms like multiple Compton scattering and synchrotron radiation (i.e. radiation which occurs when charged particles are accelerated in a curved path or orbit). Using the conservation of entropy the equation of state can be expressed by
\begin{equation}
\frac{D}{D\tau }(P_s/n_s^{\gamma _\textrm{g}})=0.\label{eq34}
\end{equation}
where $D/D\tau =(1/\alpha )\partial /\partial t+{\bf v}_s\cdot \nabla $. The full equation of state for a relativistic fluid, as measured in the fluid's rest frame, is given by \cite{fifty five,fifty six}
\begin{equation}
\varepsilon =m_sn_s+m_sn_s\left[\frac{P_s}{m_sn_s}-\frac{{\rm i}H_2^{(1)^\prime }({\rm i}m_sn_s/P_s)}{{\rm i}H_2^{(1)}({\rm i}m_sn_s/P_s)}\right]\label{eq35},
\end{equation}
where the $H_2^{(1)}(x)$ are Hankel functions.

The Rindler coordinate system, in which space is locally Cartesian, provides a good approximation to the Reissner-Nordstr\"{o}m-de Sitter geometry near the event horizon. The essential features of the horizon and the 3+1 split of spacetime are preserved without the complication of explicitly curved spatial three-geometries. In Rindler coordinates the Reissner-Nordstr\"{o}m-de Sitter metric is approximated by
\begin{equation}
ds^2=-\alpha^2 dt^2+dx^2+dy^2+dz^2,\label{eq36}
\end{equation}
where
\begin{equation}
x=r_+\left(\theta -\frac{\pi }{2}\right),\quad y=r_+\varphi ,\quad z=2r_+\alpha \label{eq37}.
\end{equation}
The standard lapse function simplifies in Rindler coordinates to $\alpha =z/2r_+$, where $r_+$ is the event horizon of the black hole.

\section{Wave Propagation in Radial Direction}\label{sec4}

We consider one-dimensional wave propagation in the radial $z$ direction. Introducing the complex variables
\begin{eqnarray}
&&v_{sz}(z,t)=u_s(z,t),\qquad v_s(z,t)=v_{sx}(z,t)+{\rm i}v_{sy}(z,t),\nonumber\\
&&B(z,t)=B_x(z,t)+{\rm i}B_y(z,t),\qquad
E(z,t)=E_x(z,t)+{\rm i}E_y(z,t)\label{eq38},
\end{eqnarray}
and defining
\begin{eqnarray}
&&v_{sx}=\frac{1}{2}(v_s+v_s^\ast),\quad v_{sy}=\frac{1}{2\textrm{i}}(v_s-v_s^\ast),\nonumber\\
&&B_x=\frac{1}{2}(B+B^\ast),\quad B_y=\frac{1}{2\textrm{i}}(B-B^\ast)\label{eq39},
\end{eqnarray}
we obtain
\begin{equation}
v_{sx}B_y-v_{sy}B_x=\frac{\rm i}{2}(v_sB^\ast -v_s^\ast B),\label{eq40}
\end{equation}
and similarly for the transverse components of $\textbf{E}$, where the $\ast $ denotes the complex conjugate. The gravitational acceleration becomes
\begin{eqnarray}
a=-\frac{\partial}{\partial z}{(\ln \alpha)}=-\frac{1}{\alpha}\frac{\partial\alpha}{\partial z}.\label{eq41}
\end{eqnarray}

The equation of continuity (\ref{eq20}) and Poisson's equation (\ref{eq24}) are respectively written as
\begin{equation}
\frac{\partial }{\partial t}(\gamma _sn_s)+\frac{\partial }{\partial z}(\alpha \gamma _sn_su_s)=0,\label{eq42}
\end{equation}
and
\begin{equation}
\frac{\partial E_z}{\partial z}=4\pi (q_1n_1\gamma _1+q_2n_2\gamma _2).\label{eq43}
\end{equation}
From the ${\bf e}_{\hat x}$ and ${\bf e}_{\hat y}$ components of (\ref{eq25}) and (\ref{eq26}), one could derive
\begin{eqnarray}
\frac{1}{\alpha }\frac{\partial B}{\partial t}&=&-{\rm i}\left(\frac{\partial }{\partial z}-a\right)E,\label{eq44}\\
{\rm i}\frac{\partial E}{\partial t}&=&-\alpha \left(\frac{\partial }{\partial z}-a\right)B-{\rm i}4\pi e\alpha (\gamma _2n_2v_2-\gamma _1n_1v_1)\label{eq45}.
\end{eqnarray}
Differentiating (\ref{eq45}) with respect to $t$ and using (\ref{eq44}), we obtain
\begin{eqnarray}
\left(\alpha ^2\frac{\partial ^2}{\partial z^2}+3\alpha \frac{\partial \alpha}{\partial z}\frac{\partial }{\partial z}-\frac{\partial ^2}{\partial t^2}+\left(\frac{\partial \alpha }{\partial z}\right)^2\right)E=4\pi e\alpha \frac{\partial }{\partial t}(n_2\gamma _2v_2-n_1\gamma _1v_1)\label{eq46}.
\end{eqnarray}
From the ${\bf e}_{\hat x}$ and ${\bf e}_{\hat y}$ components of (\ref{eq32}), the transverse component of the momentum conservation equation is obtained in the form
\begin{eqnarray}
\rho _s\frac{Dv_s}{D\tau }=q_sn_s\gamma _s(E-{\rm i}v_sB_z+{\rm i}u_sB)-u_sv_s\rho _sa-v_s\left(q_sn_s\gamma _s{\bf E}\cdot {\bf v}_s+\frac{1}{\alpha }\frac{\partial P_s}{\partial t}\right),\label{eq47}
\end{eqnarray}
where
\[
{\bf E}\cdot {\bf v}_s=\frac{1}{2}(Ev_s^\ast +E^\ast v_s)+E_zu_s,
\]
and $\rho _s$ is the total energy density defined by \begin{equation}
\rho _s=\gamma _s^2(\varepsilon _s+P_s)=\gamma _s^2(m_sn_s+\Gamma _\textrm{g}P_s),\label{eq48}
\end{equation}
with $\Gamma _\textrm{g}=\gamma _\textrm{g}/(\gamma _\textrm{g}-1)$.

\section{Linearization}\label{sec5}

We use perturbation method to linearize the equations derived in the preceding section. We choose an applied magnetic field to lie along the radial ${\bf e}_{\hat z}$ direction and introduce the quantities
\begin{eqnarray}
u_s(z,t)&=&u_{0s}(z)+\delta u_s(z,t),\quad v_s(z,t)=\delta v_s(z,t),\nonumber\\
n_s(z,t)&=&n_{0s}(z)+\delta n_s(z,t),\quad P_s(z,t)=P_{0s}(z)+\delta P_s(z,t),\nonumber\\
\rho _s(z,t)&=&\rho _{0s}(z)+\delta \rho _s(z,t),\quad {\bf E}(z,t)=\delta {\bf E}(z,t),\nonumber\\
{\bf B}_z(z,t)&=&{\bf B}_0(z)+\delta {\bf B}_z(z,t),\quad {\bf B}(z,t)=\delta {\bf B}(z,t).\label{eq49}
\end{eqnarray}
The relativistic Lorentz factor is also linearized such that
\begin{equation}
\gamma _s=\gamma _{0s}+\delta \gamma _s,\quad\mbox{where}\quad \gamma _{0s}=\left(1-{\bf u}_{0s}^2\right)^{-\frac{1}{2}},\quad\delta \gamma _s=\gamma _{0s}^3{\bf u}_{0s}\cdot \delta {\bf u}_s\label{eq50}.
\end{equation}
The unperturbed radial velocity near the event horizon for each species as measured by a FIDO along ${\bf e}_{\hat z}$ is assumed to be the freefall velocity so that
\begin{eqnarray}
u_{0s}(z)=v_{\rm ff}(z)=[1-\alpha ^2(z)]^{\frac{1}{2}}\label{eq51},
\end{eqnarray}
which, with (\ref{eq3}) and (\ref{eq17}), gives
\begin{eqnarray}
v_{\rm ff}(z)=\left({\frac{r_+}{r}}\right)^{\frac{1}{2}}\chi,\label{eq52}
\end{eqnarray}
where
\begin{eqnarray}
\chi=\left[1+\frac{r}{2M}\left(\frac{r^2 }{\ell^2}-\frac{q^2}{r^2}\right) \right]^{\frac{1}{2}}\left[1+\frac{r_+}{2M}\left(\frac{r_+^2 }{\ell^2}-\frac{q^2}{r_+^2}\right) \right]^{-\frac{1}{2}}\label{eq53}.
\end{eqnarray}
For the plasma falling at the black hole event horizon, $v_{\textrm{ff}}=1$ and at cosmological horizon, $v_{\textrm{ff}}=(r_+/r_c)<1$, while for the plasma inside the cosmological horizon and falling at the black hole horizon, we have $1 < \chi<(r_+/r_c)^{1/2}$. It follows, from the continuity equation (\ref{eq42}), that $r^2\alpha \gamma _{0s}n_{0s}u_{0s}=\mbox{const.}=r_+^2\alpha _+\gamma _+n_+u_+$, where the values with a subscript \lq \lq $+$\rq \rq are the limiting values at the event horizon. The freefall velocity at the horizon becomes unity so that $u_+=1$. Since $u_{0s}=v_{\rm ff}$, $\gamma _{0s}=1/\alpha $; and hence $\alpha \gamma _{0s}=\alpha _+\gamma _+=1$. Also, because $v_{\rm ff}=\chi(r_+/r)^{1/2}$, the number density for each species can be written as follows:
\begin{equation}
n_{0s}(z)=\frac{1}{\chi ^4}n_{+s}v_{\rm ff}^3(z)\label{eq54}.
\end{equation}
The equation of state (\ref{eq34}) and (\ref{eq54}) lead to write the unperturbed pressure, in terms of the freefall velocity, as follows:
\begin{equation}
P_{0s}(z)=\frac{1}{\chi ^{4\gamma_g}}P_{+s}v_{\rm ff}^{3\gamma _{\textrm{g}}}(z)\label{eq55}.
\end{equation}
Since $P_{0s}=k_Bn_{0s}T_{0s}$, then with $k_B=1$, the temperature profile is
\begin{equation}
T_{0s}(z)=\frac{1}{\chi ^{4(\gamma_{\textrm{g}}-1)}}T_{+s}v_{\rm ff}^{3(\gamma _{\textrm{g}}-1)}(z)\label{eq56}.
\end{equation}

The unperturbed magnetic field is purely in the radial direction. It does not experience effects of spatial curvature. From the flux conservation $\nabla \cdot {\bf B}_0=0$, it follows that $r^2B_0(r)=\mbox{const.}$, from which one obtains the unperturbed magnetic field, in terms of the freefall velocity as
\begin{equation}
B_0(z)=\frac{1}{\chi ^4}B_+v_{\rm ff}^4(z),\label{eq57}
\end{equation}
where $v_{\rm ff}(z)=[1-\alpha ^2(z)]^{1/2}$. Since
\begin{equation}
\frac{dv_{\rm ff}}{dz}=-\frac{\alpha }{2r_+}\frac{1}{v_{\rm ff}},\label{eq58}
\end{equation}
we have
\begin{eqnarray}
\frac{du_{0s}}{dz}&=&-\frac{\alpha }{2r_+}\frac{1}{v_{\rm ff}},\quad \frac{dB_0}{dz}=-\frac{4\alpha }{2r_+}\frac{B_0}{v_{\rm ff}^2},\nonumber\\
\frac{dn_{0s}}{dz}&=&-\frac{3\alpha }{2r_+}\frac{n_{0s}}{v_{\rm ff}^2},\quad \frac{dP_{0s}}{dz}=-\frac{3\alpha }{2r_+}\frac{\gamma _{\textrm{g}}P_{0s}}{v_{\rm ff}^2}\label{eq59}.
\end{eqnarray}

When the linearized variables from (\ref{eq49}) and (\ref{eq50}) are substituted into the continuity equation and products of perturbation terms are neglected, it results
\begin{eqnarray}
\gamma _{0s}\left(\frac{\partial }{\partial t}+u_{0s}\alpha \frac{\partial }{\partial z}+u_{0s}\frac{\partial \alpha }{\partial z}+\gamma _{0s}^2\alpha \frac{du_{0s}}{dz}\right)\delta n_s+\left(\alpha \frac{\partial }{\partial z}+\frac{\partial \alpha }{\partial z}\right)(n_{0s}\gamma _{0s}u_{0s})\nonumber\\
+n_{0s}\gamma _{0s}^3\left[u_{0s}\frac{\partial }{\partial t}+\alpha \frac{\partial }{\partial z}
+\frac{\partial \alpha }{\partial z}+\alpha \left(\frac{1}{n_{0s}}\frac{dn_{0s}}{dz}+3\gamma _{0s}^2u_{0s}\frac{du_{0s}}{dz}\right)\right]\delta u_s=0\label{eq60}.
\end{eqnarray}
The conservation of entropy, (\ref{eq34}), when linearized, gives
\begin{equation}
\delta P_s=\frac{\gamma _{\textrm{g}}P_{0s}}{n_{0s}}\delta n_s,\label{eq61}
\end{equation}
and hence from the total energy density, (\ref{eq48}), it is possible to write
\begin{equation}
\delta \rho _s=\frac{\rho _{0s}}{n_{0s}}\left(1+\frac{\gamma _{0s}^2\gamma _\textrm{g}P_{0s}}{\rho _{0s}}\right)\delta n_s+2u_{0s}\gamma _{0s}^2\rho _{0s}\delta u_s,\label{eq62}
\end{equation}
where $\rho _{0s}=\gamma _{0s}^2(m_sn_{0s}+\Gamma _\textrm{g}P_{0s})$. When the transverse part of the momentum conservation equation is linearized, differentiated with respect to $t$, and substituted from (\ref{eq44}), it gives
\begin{eqnarray}
\left(\alpha u_{0s}\frac{\partial }{\partial z}+\frac{\partial }{\partial t}-u_{0s}\frac{\partial \alpha }{\partial z}+\frac{{\rm i}\alpha q_s\gamma _{0s}n_{0s}B_0}{\rho _{0s}}\right)\frac{\partial \delta v_s}{\partial t}\nonumber\\
-\frac{\alpha q_s\gamma _{0s}n_{0s}}{\rho _{0s}}\left(\alpha u_{0s}\frac{\partial }{\partial z}+\frac{\partial }{\partial t}+u_{0s}\frac{\partial \alpha }{\partial z}\right)\delta E=0\label{eq63}.
\end{eqnarray}
Poisson's equation (\ref{eq43}) and (\ref{eq46}) are linearized to obtain, respectively,
\begin{eqnarray}
&\frac{\partial \delta E_z}{\partial z}=4\pi e(n_{02}\gamma _{02}-n_{01}\gamma _{01})+4\pi e(\gamma _{02}\delta n_2-\gamma _{01}\delta n_1)&\nonumber\\
&+4\pi e(n_{02}u_{02}\gamma _{02}^3\delta u_2-n_{01}u_{01}\gamma _{01}^3\delta u_1),&\label{eq64}
\end{eqnarray}
\begin{eqnarray}
\left(\alpha ^2\frac{\partial ^2}{\partial z^2}+3\alpha \frac{\partial \alpha }{\partial z}\frac{\partial }{\partial z}-\frac{\partial ^2}{\partial t^2}+\left(\frac{\partial \alpha }{\partial z}\right)^2\right)\delta E=4\pi e\alpha \left(n_{02}\gamma _{02}\frac{\partial \delta v_2}{\partial t}-n_{01}\gamma _{01}\frac{\partial \delta v_1}{\partial t}\right)\label{eq65}.
\end{eqnarray}

\section{Dispersion Relation}\label{sec6}
We investigate the information which may be obtained from the linearized equations derived above by restricting our consideration to effects on a local scale for which the distance from the horizon does not vary significantly. We use a local (or mean-field) approximation for the lapse function $\alpha(z)$ and hence for the equilibrium fields and fluid quantities. Since the plasma is situated relatively close to the horizon, a relatively small change in distance $z$ will make a significant difference to the magnitude of $\alpha $. It is thus important to choose a sufficiently small range in $z$ for which $\alpha(z)$ does not vary much. We consider thin layers in the ${\bf e}_{\hat z}$ direction, each layer with its own $\alpha_0$, where $\alpha_0$ is some mean value of $\alpha$ within a particular layer. Then a more complete picture can be built up by considering a large number of layers within a chosen range of $\alpha _0$ values.

The local approximation imposes a restriction on the wavelength and hence on the wave number $k$. It is assumed that the wavelength is small compared with the range over which the equilibrium quantities change significantly. Hence, the wavelength must be smaller in magnitude than the scale of the gradient of the lapse function $\alpha $, i.e.,
\[
\lambda <\left(\frac{\partial \alpha }{\partial z}\right)^{-1}=2
r_{+} \simeq\frac{1}{2\xi}\left[1+\sqrt{1-\frac{2\xi}{1+\zeta} \frac{q^2}{M^2}}\right]5.896\times 10^5{\rm cm},
\]
or, equivalently,
\[
k>\frac{2\pi }{2r_+}\simeq 2\xi\left[1+\sqrt{1-\frac{2\xi}{1+\zeta}\frac{q^2}{M^2}}\right]^{-1} 1.067\times 10^{-5}{\rm cm}^{-1},\quad\mbox{with}\quad 1\leq\frac{2\xi}{1+\zeta}\leq\frac{M^2}{q^2},
\]
for a black hole of mass $\sim 1M_\odot$. The value for $2\xi /(1+\zeta)=1$ corresponds to the RN black hole, while the value for $2\xi /(1+\zeta)=M^2/q^2$ corresponds to the extreme RNdS black hole.

Since the hydrodynamical approach used in the work is essentially a bulk, fluid approach, the microscopic behavior of the two-fluid plasma is treated in a somewhat approximate manner via the equation of state. Then the results are really only strictly valid in the long wavelength limit. However, the restriction, imposed by the local approximation, on the wavelength is not too severe and permits the consideration of intermediate to long wavelengths so that the small $k$ limit is still valid.

The unperturbed fields and fluid quantities are not assumed to be constant with respect to $\alpha(z)$. Then, the derivatives of these quantities can be evaluated at each layer for a given $\alpha_0$. In the local approximation, $\alpha\simeq\alpha_0$ is valid within a particular layer. Hence, the unperturbed fields, fluid quantities and their derivatives take on their corresponding \lq \lq mean-field\rq \rq  values for a given $\alpha_0$. The coefficients in (\ref{eq60}), (\ref{eq63}), and (\ref{eq64}) become constants within each layer, evaluated at each fixed mean-field value, $\alpha=\alpha_0$. So, it is possible to Fourier transform the equations with respect to $z$, using plane-wave-type solutions for the perturbations of the form $\sim e^{i(kz-\omega t)}$ for each $\alpha _0$ layer.

When Fourier transformed, (\ref{eq44}), (\ref{eq45}), (\ref{eq63}), and (\ref{eq65}) turn out to be
\begin{eqnarray}
&&\left(k-\frac{\rm i}{2r_+\alpha _0}\right)\delta E+\frac{{\rm i}\omega }{\alpha _0}\delta B=0,\label{eq66}\\
&&\frac{{\rm i}\omega }{\alpha _0}\delta E=\left(k-\frac{\rm i}{2r_+\alpha _0}\right)\delta B+4\pi e(\gamma _{02}n_{02}\delta v_2-\gamma _{01}n_{01}\delta v_1)\label{eq67},
\end{eqnarray}
\begin{eqnarray}
\omega \left(\alpha _0ku_{0s}-\omega +\frac{{\rm i}u_{0s}}{2r_+}+\frac{\alpha _0q_s\gamma _{0s}n_{0s}B_0}{\rho _{0s}}\right)\delta v_s\nonumber\\
-{\rm i}\alpha _0\frac{q_s\gamma _{0s}n_{0s}}{\rho _{0s}}\left(\alpha _0ku_{0s}-\omega -\frac{{\rm i}u_{0s}}{2r_+}\right)\delta E=0\label{eq68},
\end{eqnarray}
and
\begin{equation}
\delta E=\frac{{{\rm i}4\pi e\alpha _0\omega (n_{02}\gamma _{02}\delta v_2-n_{01}\gamma _{01}}\delta v_1)}{\alpha _0k(\alpha _0k-{\rm i}3/(2r_+))-\omega ^2-1/(2r_+)^2}\label{eq69}.
\end{equation}
The dispersion relation for the transverse electromagnetic wave modes may be put in the form
\begin{eqnarray}
&&\left[K_\pm \left(K_\pm \pm \frac{\rm i}{2r_+}\right)-\omega ^2+\frac{1}{(2r_+)^2}\right]\nonumber\\
&&=\alpha _0^2\left\{\frac{\omega _{p1}^2(\omega -u_{01}K_\pm )}{(u_{01}K_\mp -\omega -\alpha _0\omega _{c1})}+\frac{\omega _{p2}^2(\omega -u_{02}K_\pm )}{(u_{02}K_\mp -\omega +\alpha _0\omega _{c2})}\right\}\label{eq70}
\end{eqnarray}
for either electron-positron or electron-ion plasma, where $K_{\pm }=\alpha_0 k\pm \textrm{i}/2r_+$, $\omega _{cs}=e^2\gamma _{0s}n_{0s}B_0/\rho_{0s}$ and $\omega _{ps}=\sqrt{4\pi e^2\gamma _{0s}^2n_{0s}^2/\rho _{0s}}$. Like the plasma frequency $\omega _{ps}$, the cyclotron frequency $\omega _{cs}$ is frame independent. The fluid quantities are measured in the fluid frame, but the field $B_0$ is measured in the FIDO frame. Hence, the factors of $\gamma_{0s}$ do not cancel out explicitly. The transformation $B_0\rightarrow\gamma_{0s}B_0$ boosts the fluid frame for either fluid and thereby cancels the $\gamma_{0s}$ factors.  The \lq \lq $+$\rq \rq and \lq \lq $-$\rq \rq  denote the left ($L$) and right ($R$) modes, respectively. The dispersion relation for the $L$ mode is obtained by taking the complex conjugate of the dispersion relation for the $R$ mode. The two modes have the same dispersion relation in the special relativistic case.

\begin{figure}[h]\label{fig1}
\begin{center}
\includegraphics[scale=0.38]{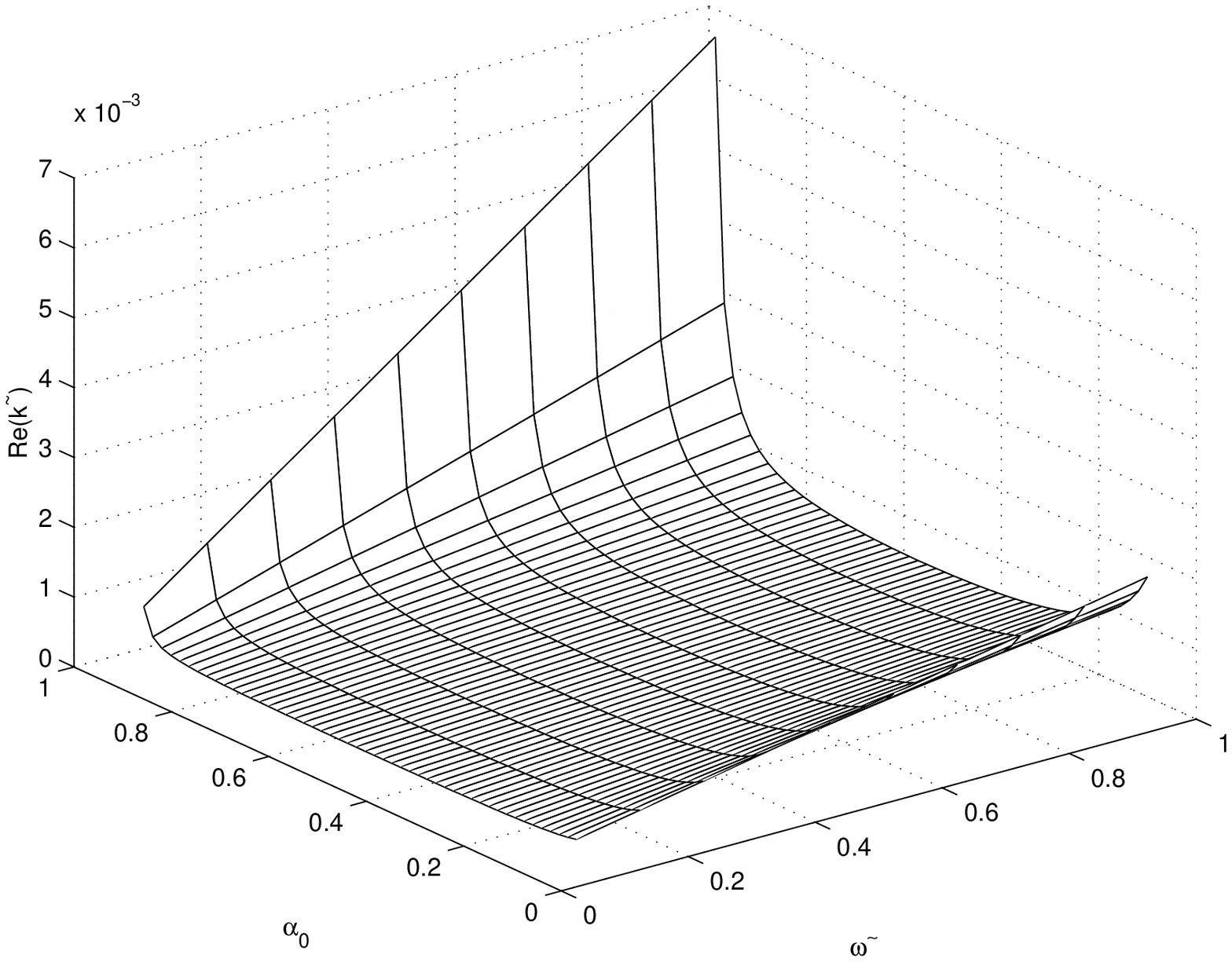}
\includegraphics[scale=0.38]{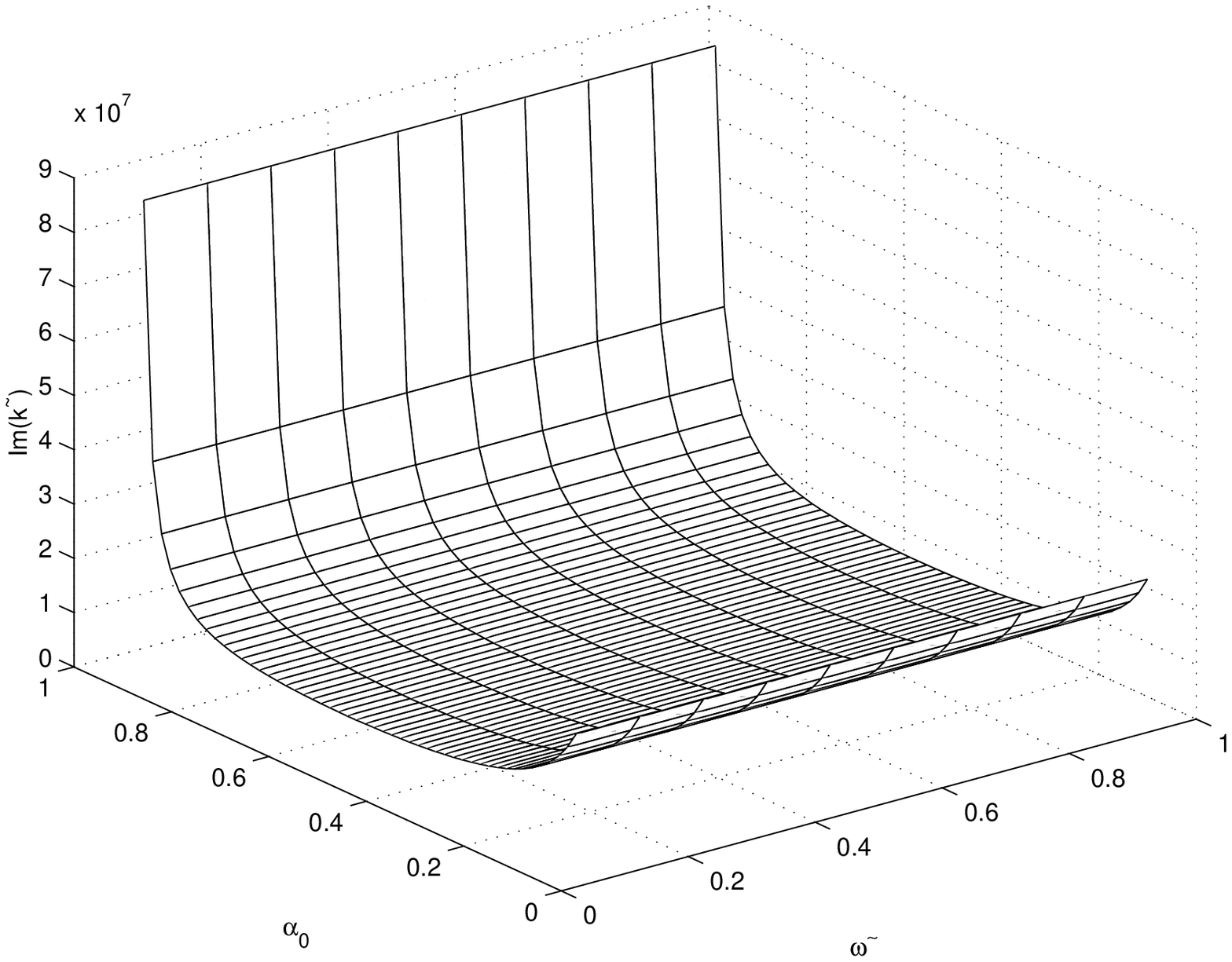}
\end{center}
\caption{\textit{Left}: Real part of Alfv\'en damped mode for the electron-positron plasma. \textit{Right}: Imaginary part of the damped mode.}
\end{figure}

\begin{figure}[h]\label{fig2}
\begin{center}
\includegraphics[scale=0.38]{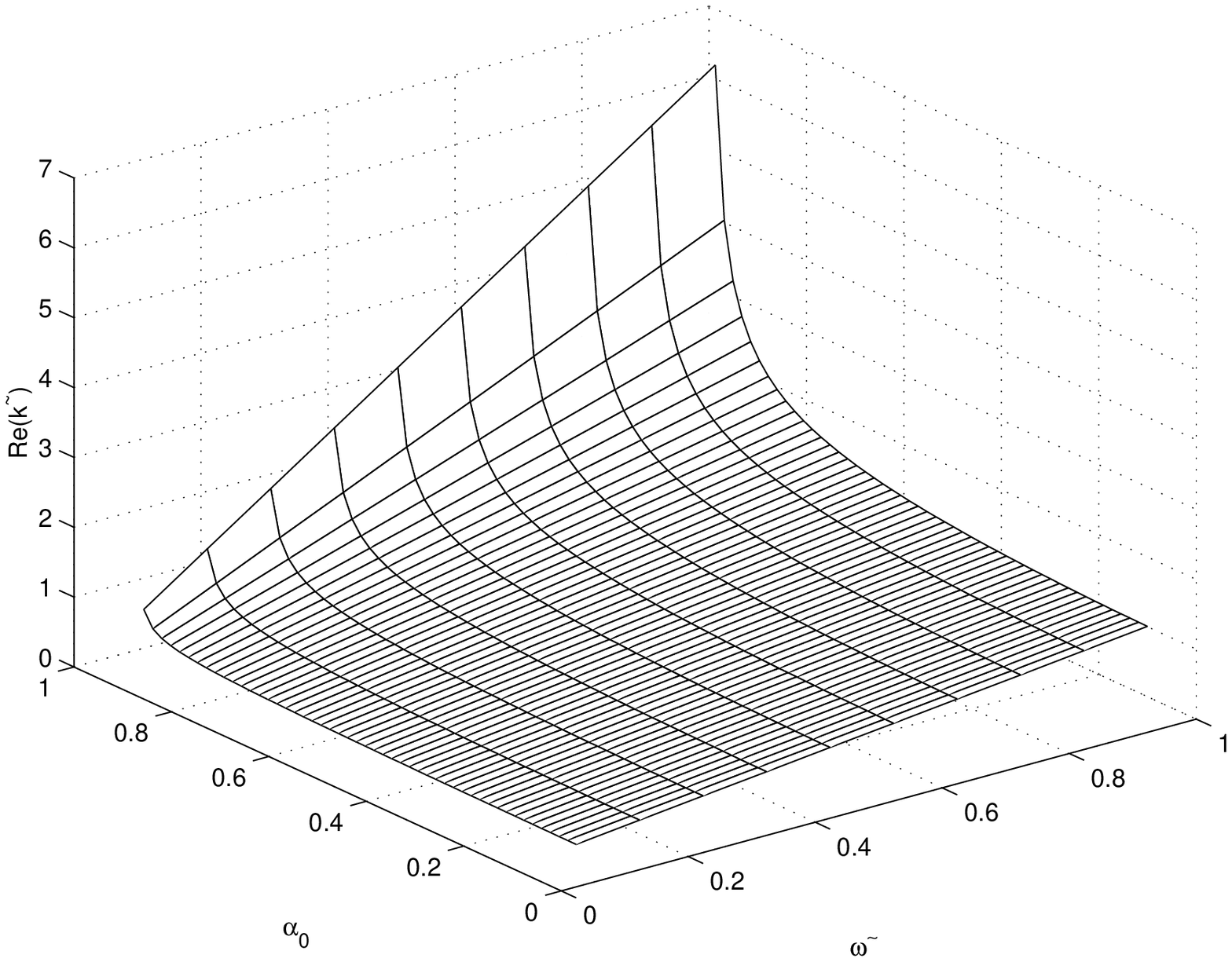}\\
\includegraphics[scale=0.38]{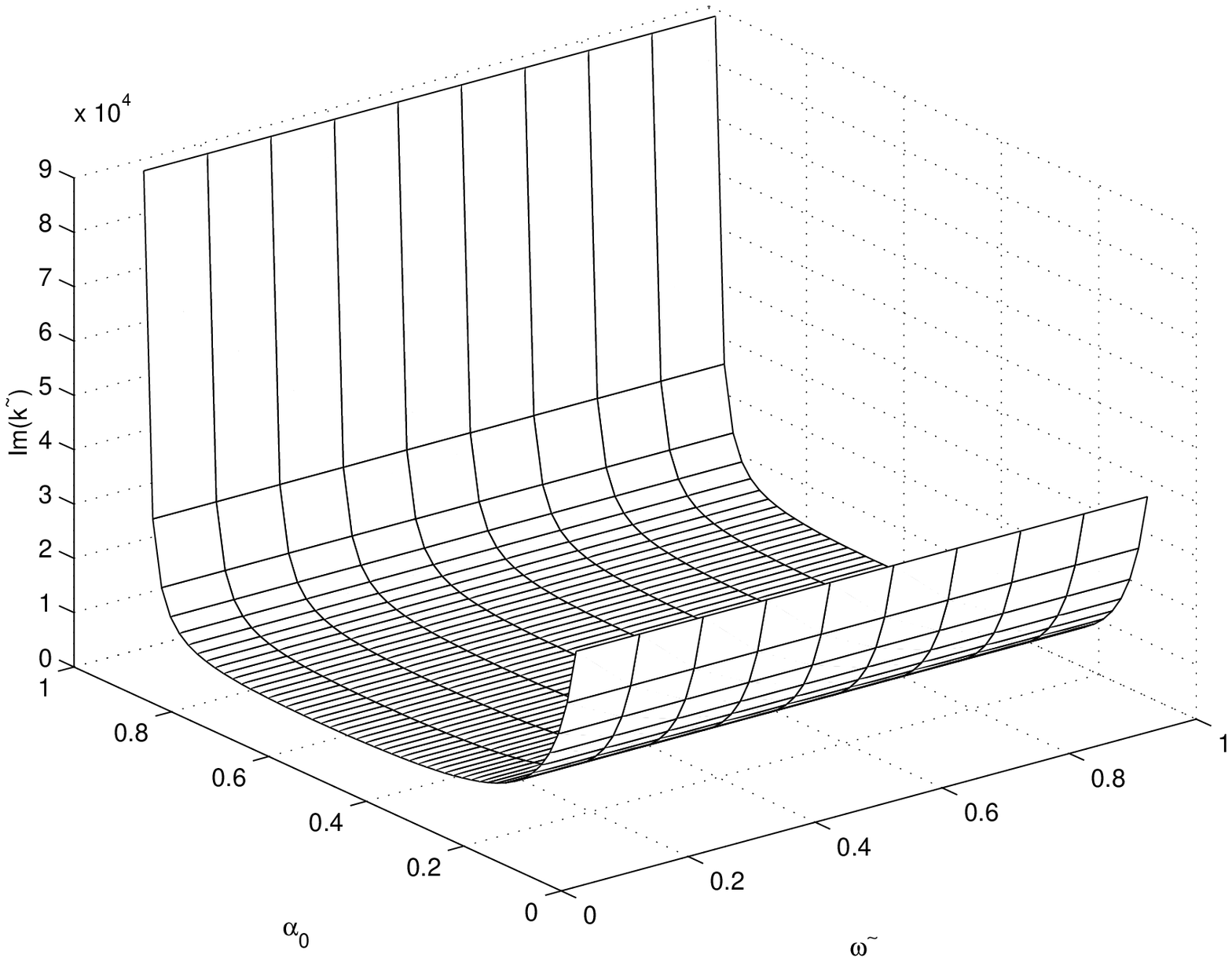}
\includegraphics[scale=0.38]{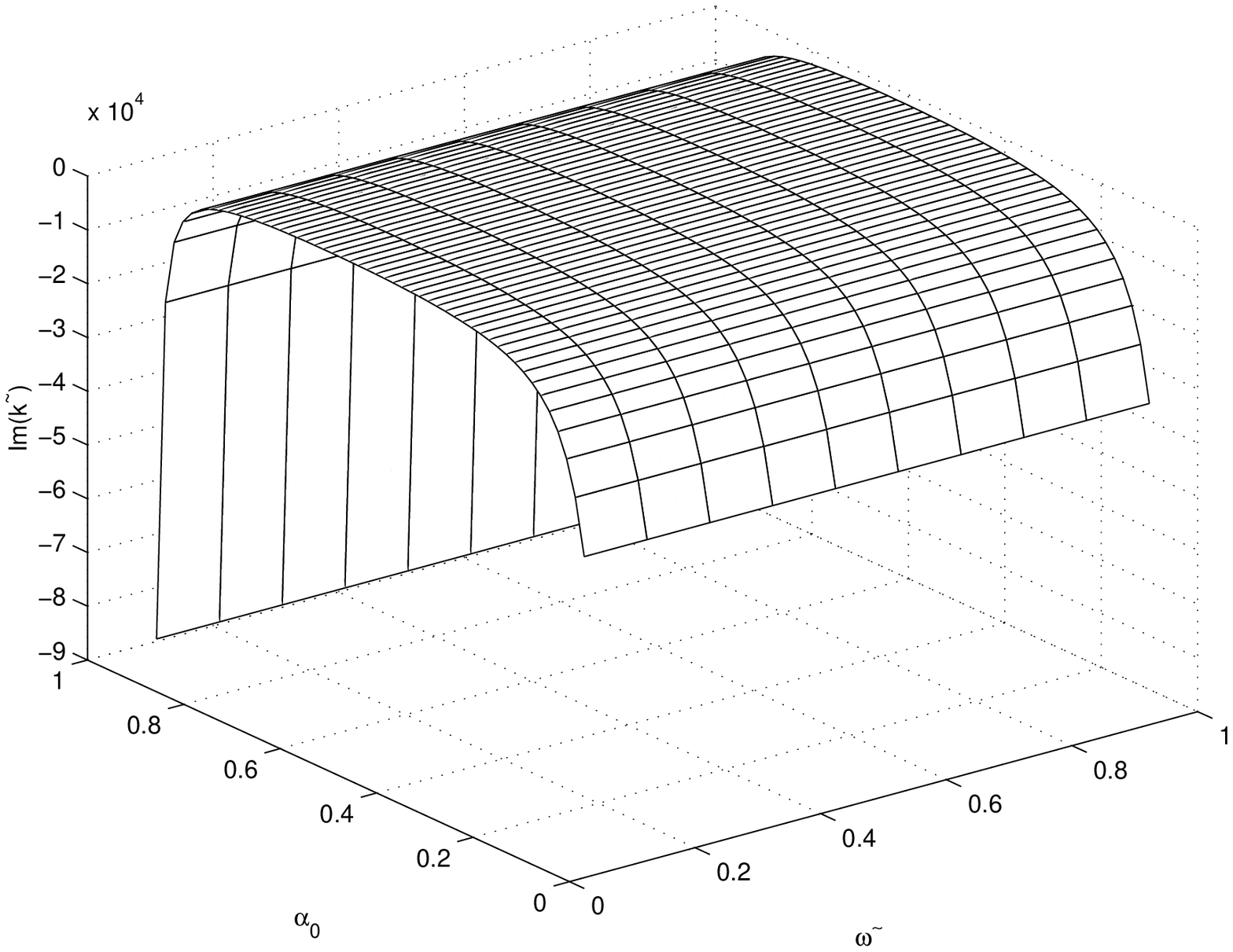}
\end{center}
\caption{\textbf{Top}: Real part of the complex conjugate pair of Alfv\'en damping and growth modes for the electron-positron plasma. \textbf{Bottom}:\textit{ Left}: Imaginary part of the damped mode. \textit{Right}: Imaginary part of the growth mode.}
\end{figure}

\section{Numerical Solution of Modes}\label{sec7}
The dispersion relations (\ref{eq70}) are complicated enough to make any attempt at an analytical solution even in the simplest cases for the electron-positron plasma where both species are assumed to have the same equilibrium parameters. As the analytical solution is cumbersome and unprofitable, we solve the dispersion relation numerically and for this purpose we put it in the form of a matrix equation as follows:
\begin{equation}
(A-kI)X=0\label{eq71},
\end{equation}
where the eigenvalue is chosen to be the wave number $k$, the eigenvector $X$ is given by the relevant set of perturbations, and $I$ is the identity matrix.

We need to write the perturbation equations in an appropriate form. The equations are then written in terms of the following set of dimensionless variables:
\begin{eqnarray}
&&\tilde \omega =\frac{\omega }{\alpha _0\omega _\ast },\quad \tilde k=\frac{kc}{\omega _\ast },\quad k_+=\frac{1}{2r_+\omega _\ast },\nonumber\\
&&\delta \tilde u_s=\frac{\delta u_s}{u_{0s}},\quad \tilde v_s=\frac{\delta v_s}{u_{0s}},\quad \delta \tilde n_s=\frac{\delta n_s}{n_{0s}},\nonumber\\
&&\delta \tilde B=\frac{\delta B}{B_0},\quad \tilde E=\frac{\delta E}{B_0},\quad \delta \tilde E_z=\frac{\delta E_z}{B_0}.\label{eq72}
\end{eqnarray}

\begin{figure}[h]\label{fig3}
\begin{center}
\includegraphics[scale=0.38]{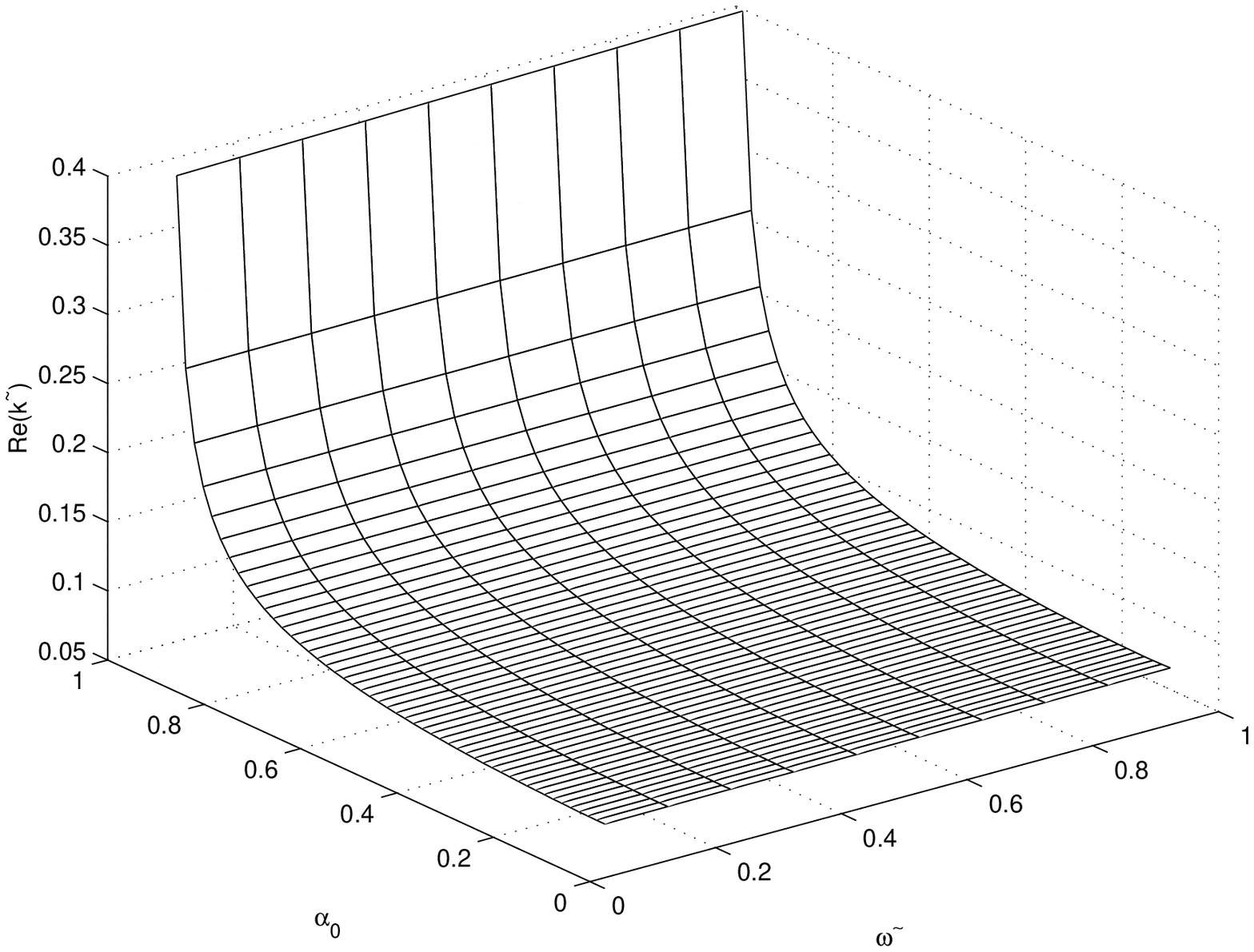}\\
\includegraphics[scale=0.38]{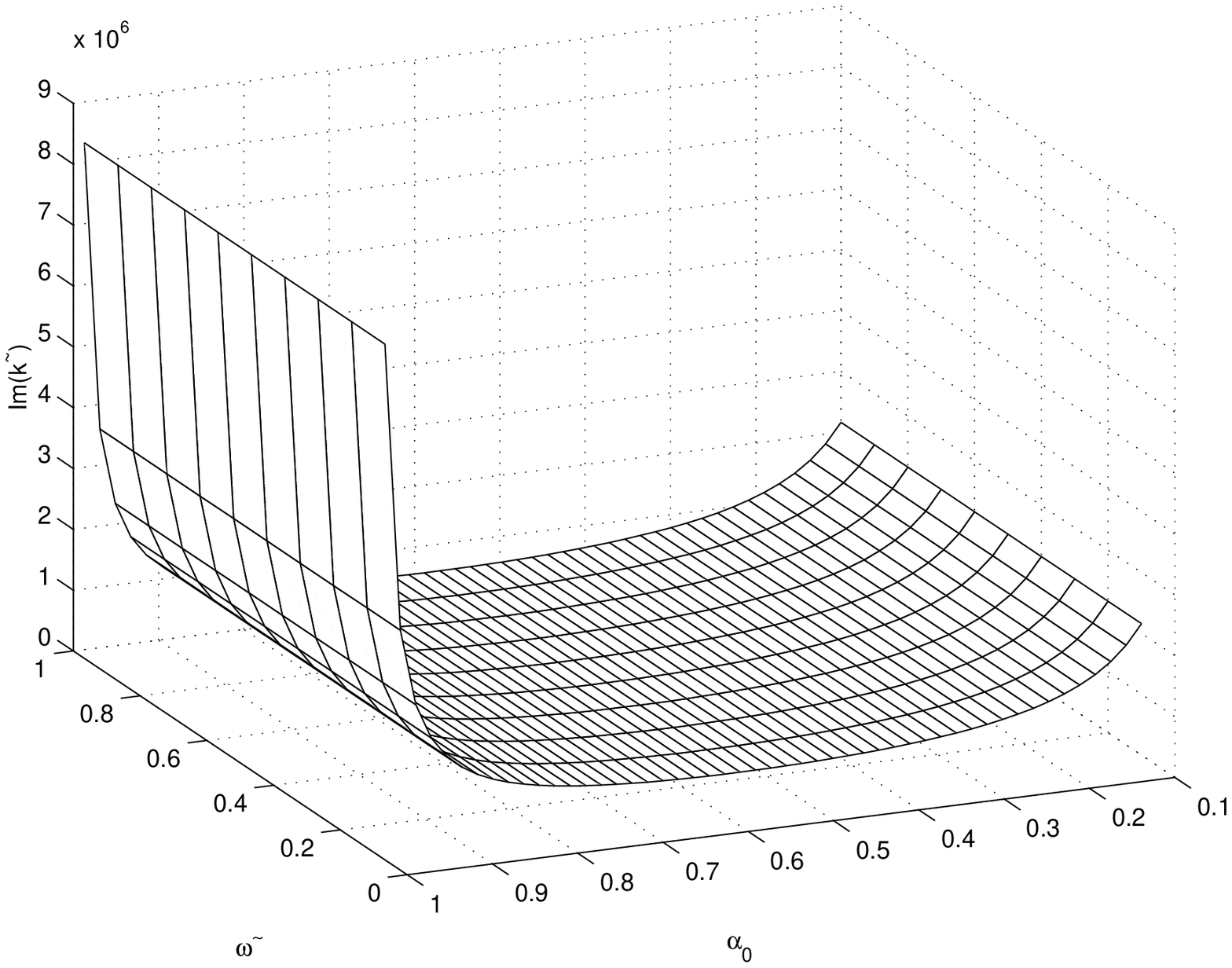}
\includegraphics[scale=0.38]{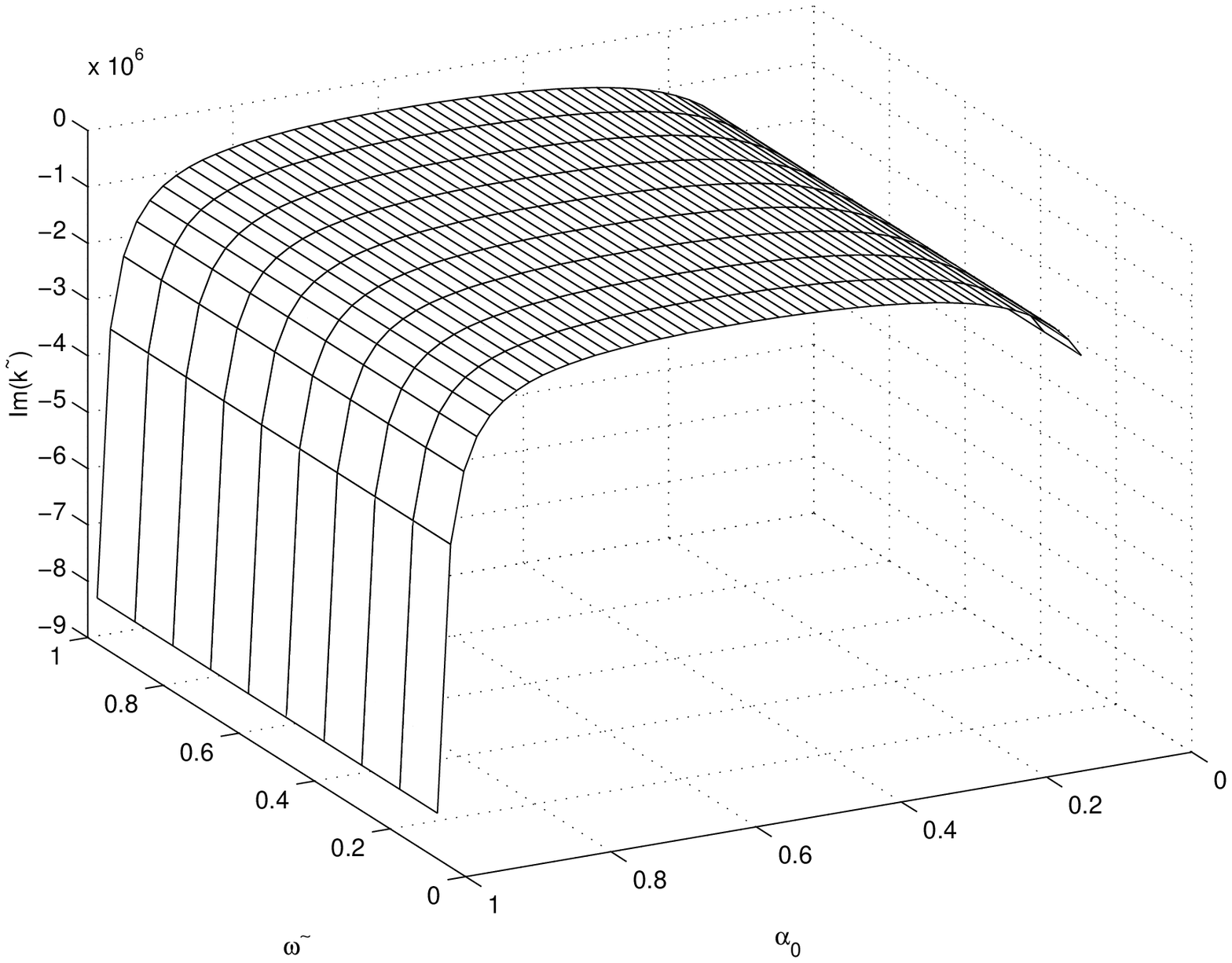}
\end{center}
\caption{\textbf{Top}: Real part of the complex conjugate pair of Alfv\'en damping and growth modes for the electron-ion plasma. \textbf{Bottom}: \textit{Left}: Imaginary part of the damped mode. \textit{Right}: Imaginary part of the growth mode.}
\end{figure}

\begin{figure}[h]\label{fig4}
\begin{center}
\includegraphics[scale=0.38]{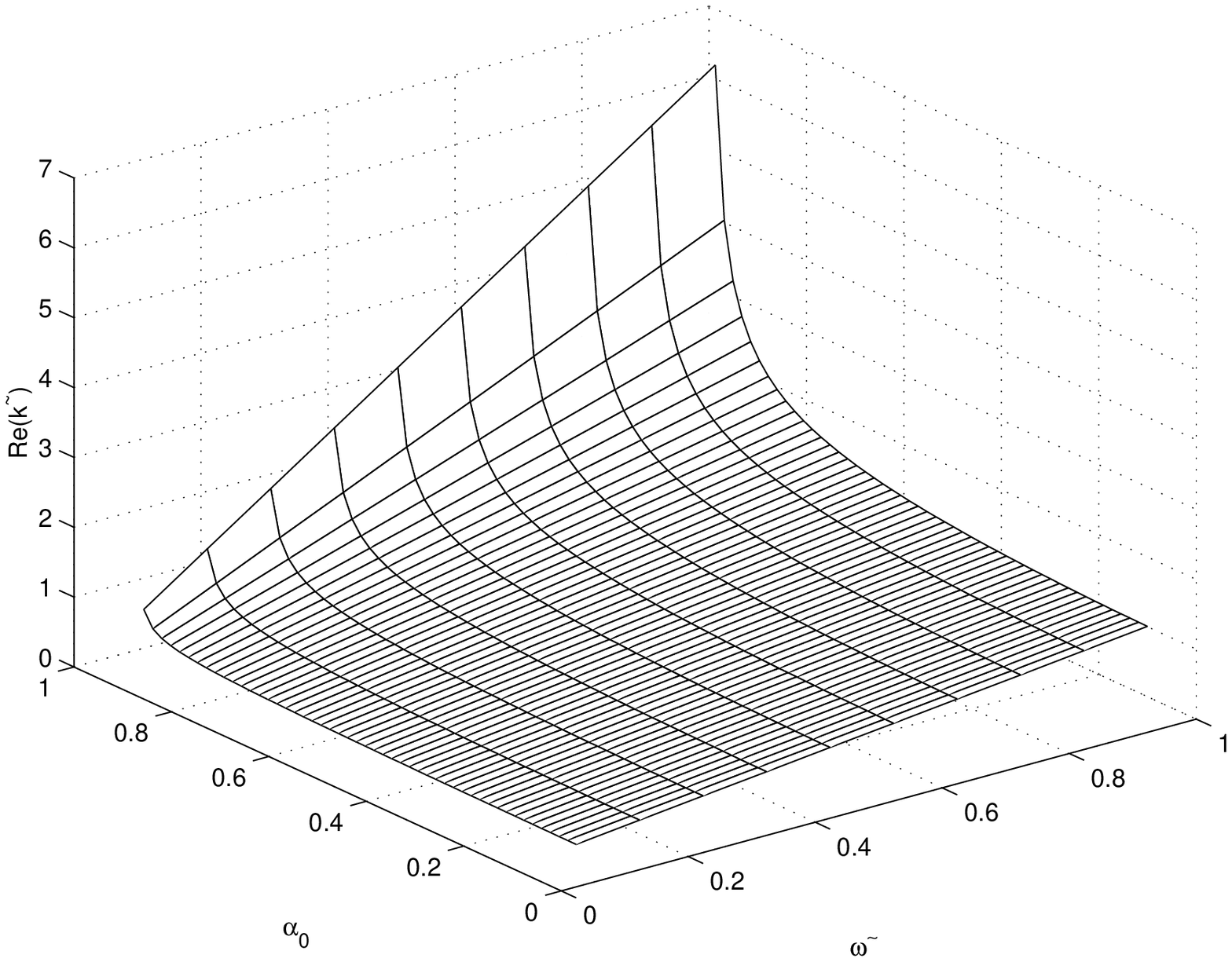}
\includegraphics[scale=0.38]{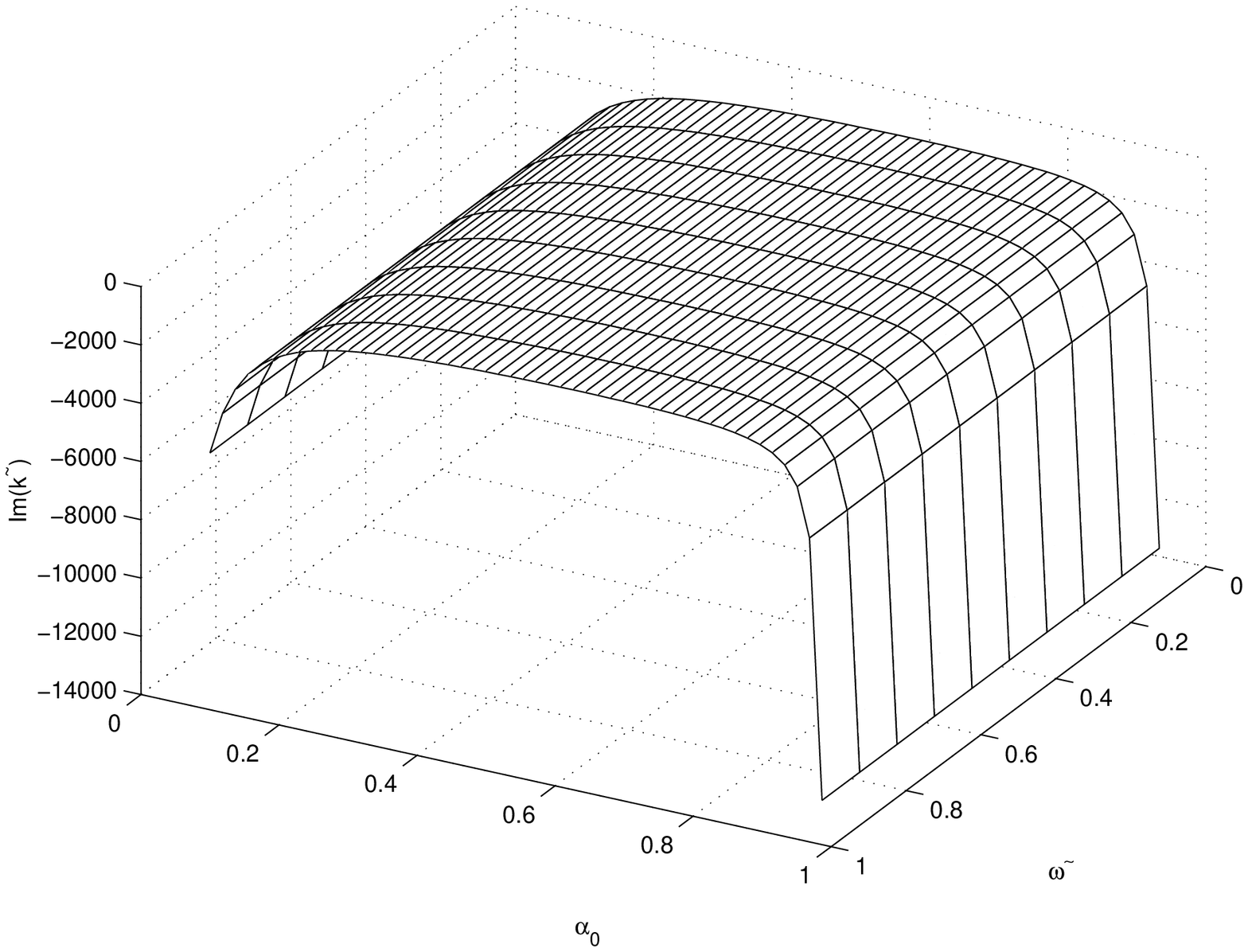}
\end{center}
\caption{\textit{Left}: Real part of Alfv\'{e}n growth mode for the electron-ion plasma. \textit{Right}: Imaginary part of the growth mode.}
\end{figure}

For an electron-positron plasma, $\omega _{p1}=\omega _{p2}$ and $\omega _{c1}=\omega _{c2}$; hence, $\omega _\ast $ is defined as
\begin{equation}
\omega _\ast =\left\{\begin{array}{rl}&\omega _c\hspace{2.5cm}\mbox{Alfv\'en modes},\\
&(2\omega _p^2+\omega _c^2)^{\frac{1}{2}}\qquad
\mbox{high frequency modes},\end{array}\right.\label{eq73}
\end{equation}
where $\omega _p=\sqrt{\omega _{p1}\omega _{p2}}$ and $\omega _c=\sqrt{\omega _{c1}\omega _{c2}}$. However, for the case of an electron-ion plasma, the plasma frequency and the cyclotron frequency are different for each fluid. The choice of $\omega _\ast $ is then a more complicated matter. For simplicity, we assume that
\begin{equation}
\omega _\ast =\left\{\begin{array}{rl}&\frac{1}{\sqrt{2}}(\omega _{c1}^2+\omega _{c2}^2)^{\frac{1}{2}}\quad \mbox{Alfv\'en modes},\\
&(\omega _{\ast 1}^2+\omega _{\ast 2}^2)^{\frac{1}{2}}\qquad
\mbox{high frequency modes},\end{array}\right.\label{eq74}
\end{equation}
where $\omega _{\ast s}^2=(2\omega _{ps}^2+\omega _{cs}^2)$. In the zero gravity limit, these values for $\omega_\ast$ reduce to the special relativistic cutoffs for electron-positron plasma. The special relativistic cutoffs are determined by the dispersion relation in view of the fact that the solutions to the dispersion relation are physical (i.e., $\textrm{Re}(k)>0$) only for certain frequency regimes. As the dispersion relation can not be handled analytically, it is difficult to determine the cutoffs in the case including gravity. Other similar combinations for $\omega _\ast$ make only a difference of a scale factor to the form of the results as $\omega _\ast$ is really only a scale factor.

\begin{figure}[h]\label{fig5}
\begin{center}
\includegraphics[scale=0.38]{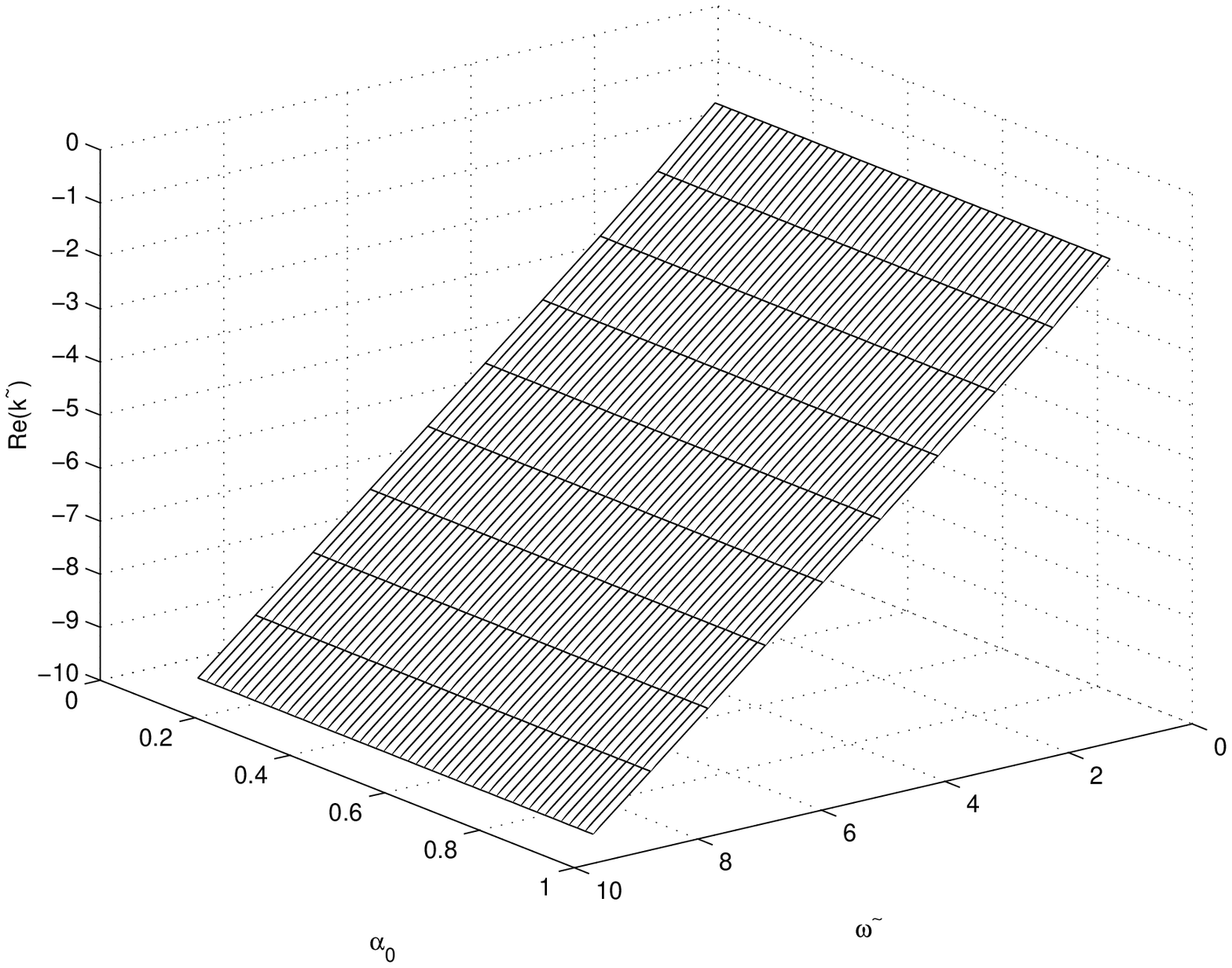}\\
\includegraphics[scale=0.38]{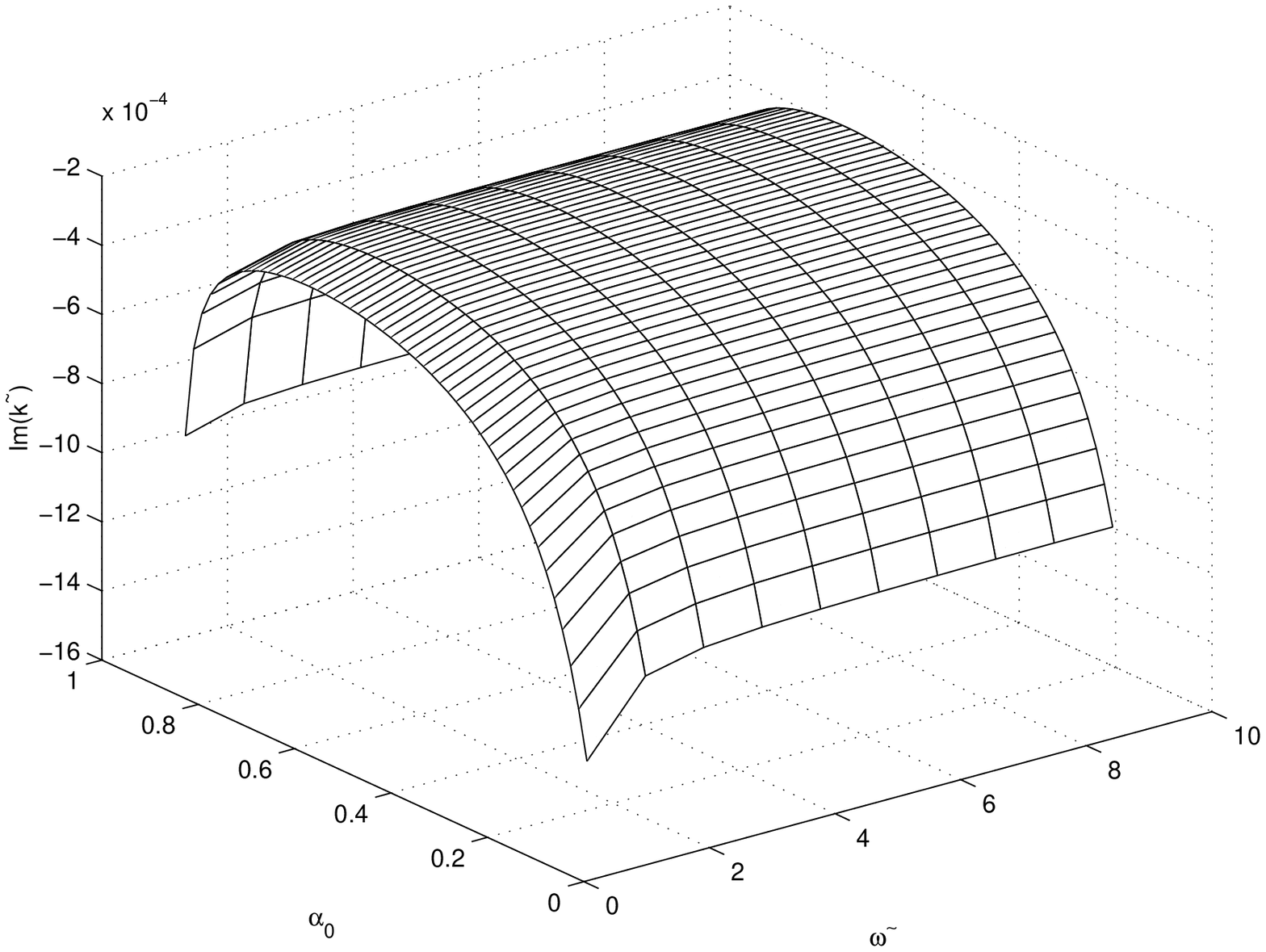}
\includegraphics[scale=0.38]{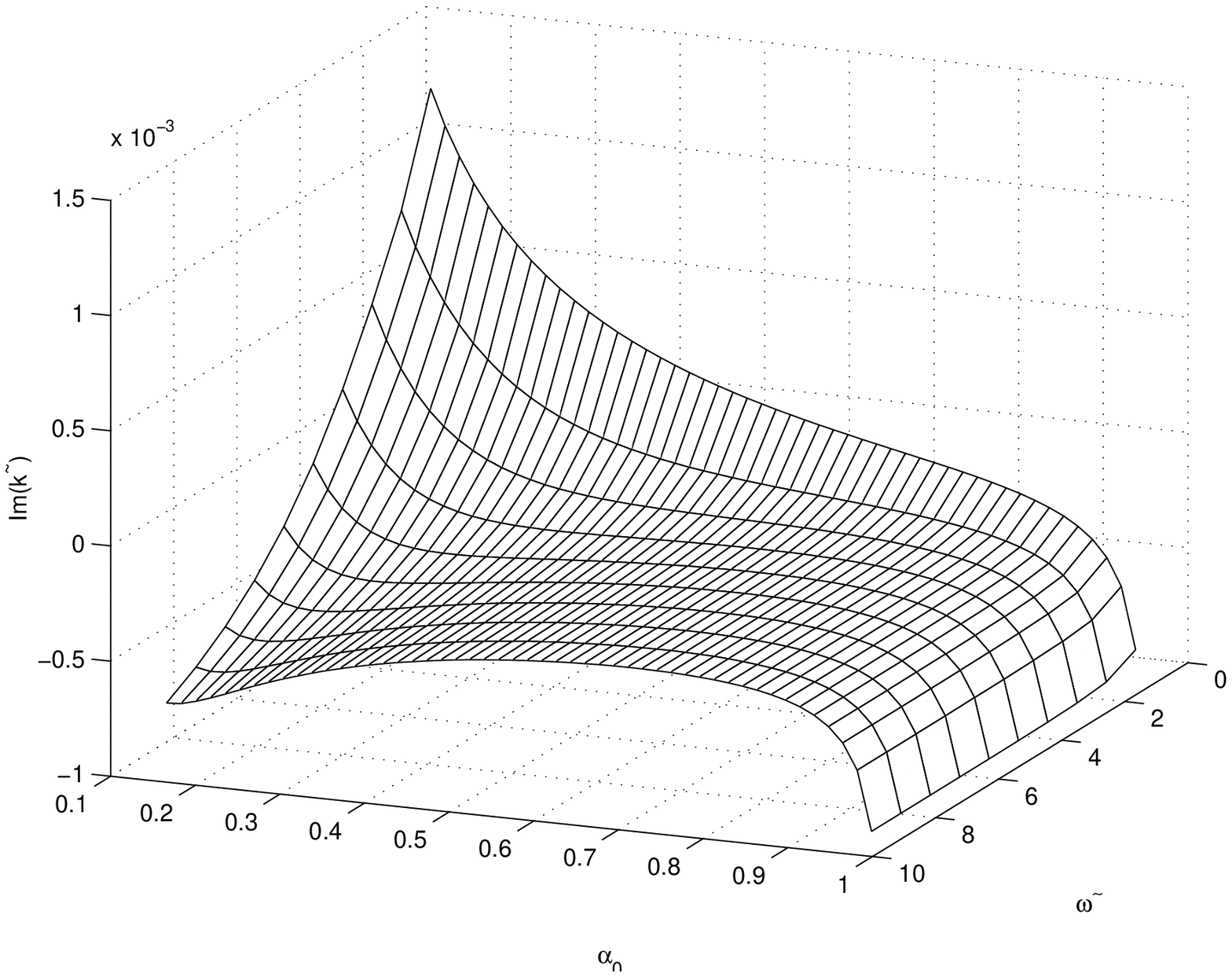}
\end{center}
\caption{\textbf{Top}: Real part of the growth mode and of the damping and growth mode for the high frequency transverse wave in the case of the electron-positron plasma. \textbf{Bottom}: \textit{Left}: Imaginary part of the growth mode. \textit{Right}: Imaginary part of the damping and growth mode.}
\end{figure}

\begin{figure}[h]\label{fig6}
\begin{center}
\includegraphics[scale=0.38]{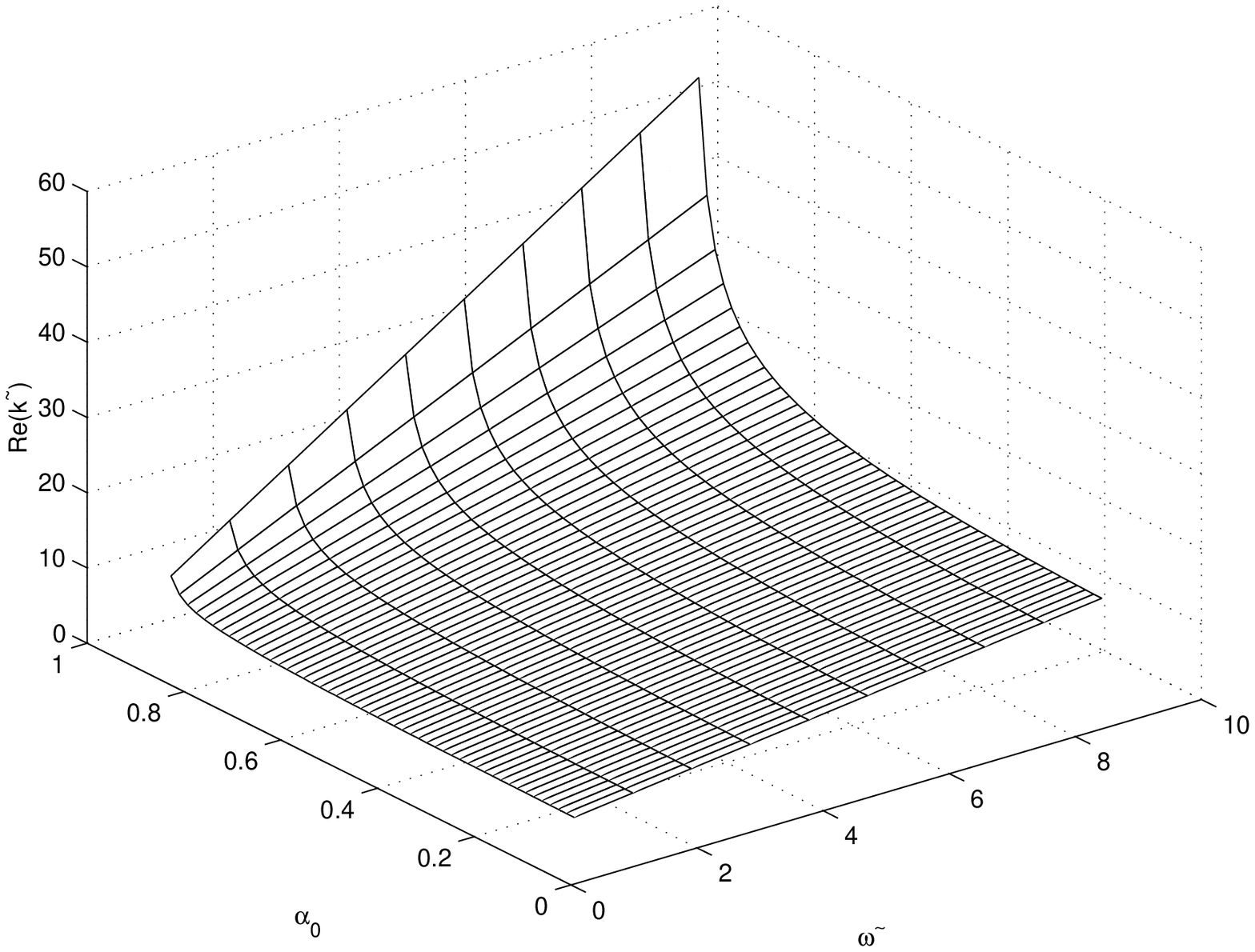}\\
\includegraphics[scale=0.38]{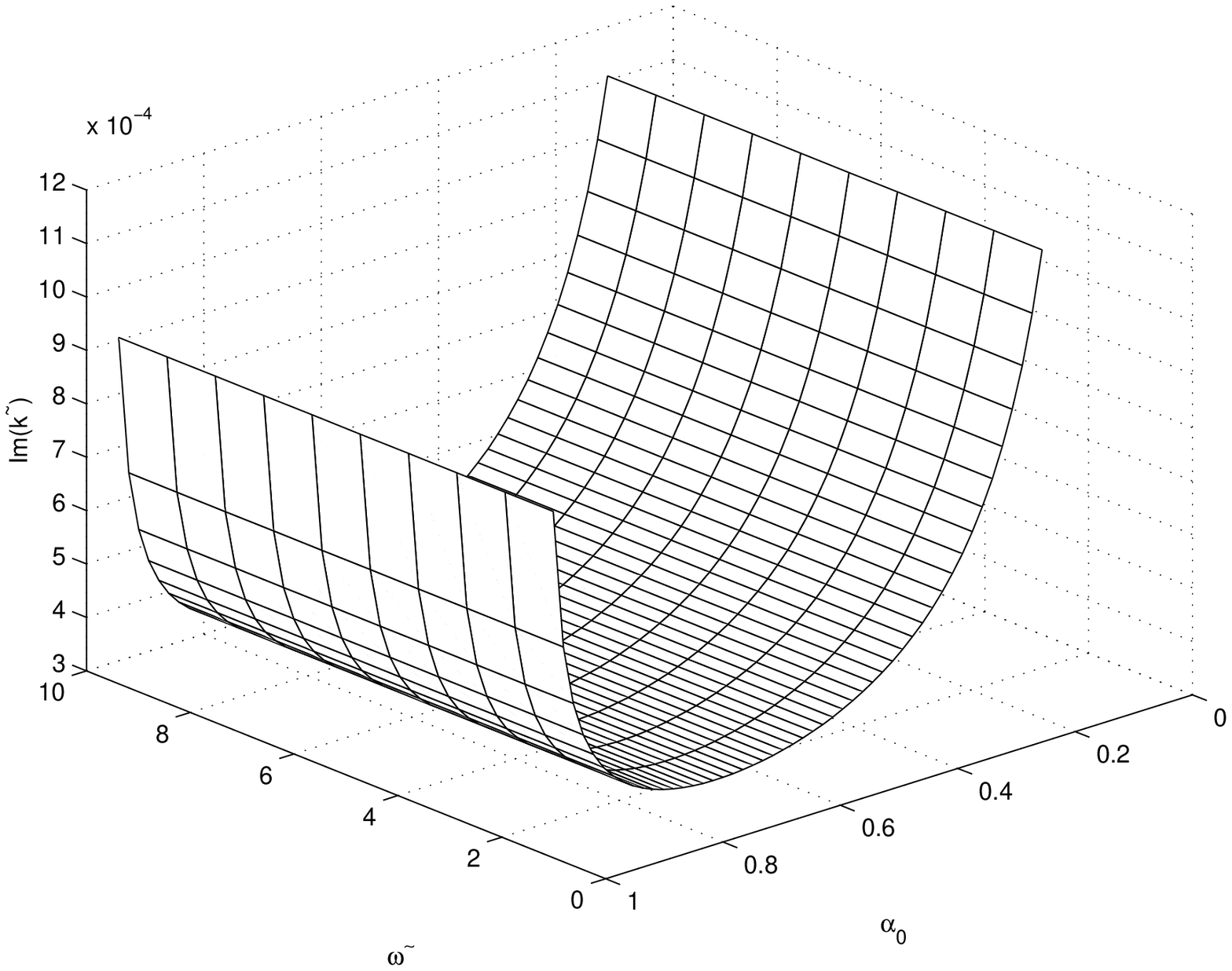}
\includegraphics[scale=0.38]{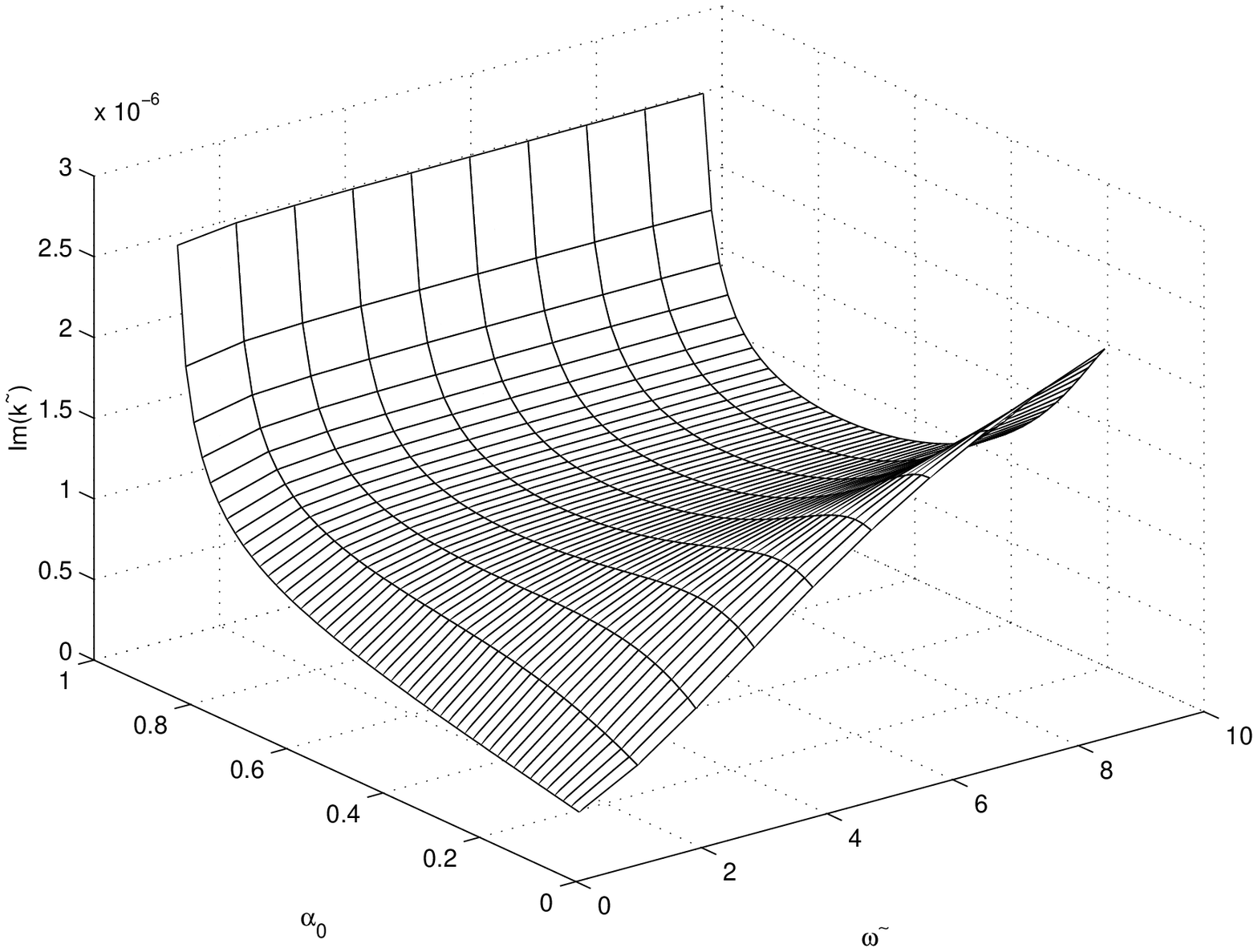}
\end{center}
\caption{\textbf{Top}: Real part of the damped mode and of the damping and growth mode for the high frequency transverse wave in the case of the electron-positron plasma. \textbf{Bottom}: \textit{Left}: Imaginary part of the damped mode. \textit{Right}: Imaginary part of the damping and growth mode.}
\end{figure}

We write the dimensionless eigenvector for the transverse set of equations in the form
\begin{equation}
\tilde X_{\rm transverse}=\left[\begin{array}{c}\delta \tilde v_1\\\delta \tilde v_2\\\delta \tilde B\\\delta \tilde E\end{array}\right]\label{eq75}.
\end{equation}
Using (\ref{eq72}), we write (\ref{eq66})--(\ref{eq68}) in the dimensionless form as follows:
\begin{equation}
\tilde k\delta \tilde E=-{\rm i}\tilde \omega \delta \tilde B+\frac{{\rm i}k_+}{\alpha _0}\delta \tilde E\label{eq76},
\end{equation}
\begin{equation}
\tilde k\delta \tilde B=u_{01}\frac{\omega _{p1}^2}{\omega _{c1}\omega _\ast }\delta \tilde v_1-u_{02}\frac{\omega _{p2}^2}{\omega _{c2}\omega _\ast }\delta \tilde v_2+\frac{{\rm i}k_+}{\alpha _0}\delta \tilde B+{\rm i}\tilde \omega \delta \tilde E\label{eq77},
\end{equation}
and
\begin{equation}
\tilde k\delta \tilde v_s=\left(\frac{\tilde \omega }{u_{0s}}-\left(\frac{q_s}{e}\right)\frac{\omega _{cs}}{u_{0s}\omega _\ast }-\frac{{\rm i}k_+}{\alpha _0}\right)\delta \tilde v_s+\left(\frac{q_s}{e}\right)\frac{\omega _{cs}}{u_{0s}\omega _\ast }\delta \tilde B-{\rm i}\left(\frac{q_s}{e}\right)\frac{\omega _{cs}}{u_{0s}\omega _\ast }\delta \tilde E\label{eq78}.
\end{equation}
These are the equations in the required form to be used as input to (\ref{eq71}).

\section{Results}\label{sec8}
We have carried out the numerical analysis by using the well known MATLAB. We have taken the values $q^2=0.5M^2$, $\chi =0.9$, and $\zeta=1.8$. We have considered both the electron-positron plasma and the electron-ion plasma. The limiting horizon values for the electron-positron plasma are taken to be
\begin{equation}
n_{+s}=10^{18}\,{\rm cm}^{-3},\quad T_{+s}=10^{10}\,{\rm K},\quad B_+=3\times 10^6\,{\rm G}. \label{eq79}
\end{equation}
For the electron-ion plasma, the ions are essentially nonrelativistic, and the limiting horizon values are chosen to be
\begin{equation}
n_{+1}=10^{18}\,{\rm cm}^{-3},\quad T_{+1}=10^{10}\,{\rm K},\quad n_{+2}=10^{15}\,{\rm cm}^{-3},\quad T_{+2}=10^{12}\,{\rm K}\label{eq80}.
\end{equation}
The equilibrium magnetic field has the same value as it has for the electron-positron case. The limiting horizon temperature for each species has been taken as derived by Colpi et al. \cite{fifty four} from studies of two temperature models of spherical accretion onto black holes. For the limiting horizon densities and the limiting horizon field, simply arbitrarily those values have been chosen which are not inconsistent with the current ideas. The gas constant and the mass of the black hole have been chosen as follows:
\begin{equation}
\gamma_{\textrm{g}}=\frac{4}{3}\quad\mbox{and}\quad M=5M_\odot \label{eq81}.
\end{equation}

\subsection{Alfv\'en Modes}\label{subsec8.1}
\subsubsection{\textit{Electron-positron Plasma}}\label{subsubsec8.1.1}

For electron-positron plasma four modes exist, two of which are growth and other two are damped. The first mode, shown in Fig. 1, is damped. The second mode, not shown here, is equal to the first mode but opposite in sign and is growth. The third and fourth modes, shown in Fig. 2, are complex conjugate pairs and show damping and growth, respectively. These two modes are similar to the corresponding Alfv\'en modes of Buzzi et al. \cite{fourty seven} but the damping and growth rates are larger by several orders of magnitude compared with those modes. These four modes coalesce with two modes of Buzzi et al. \cite{fourty seven} for the Schwarzschild case and with a single mode for the ultra relativistic electron-positron plasma \cite{fifty three}. Since we are using the convention $e^{{\rm i}kz}=e^{{\rm i}[{\rm Re}(k)+{\rm iIm}(k)]z}$, the damping corresponds to $ {\rm Im(\tilde{k})}>0 $ and growth to $\rm  Im(k)<0 $.

\begin{figure}[h]\label{fig7}
\begin{center}
\includegraphics[scale=0.38]{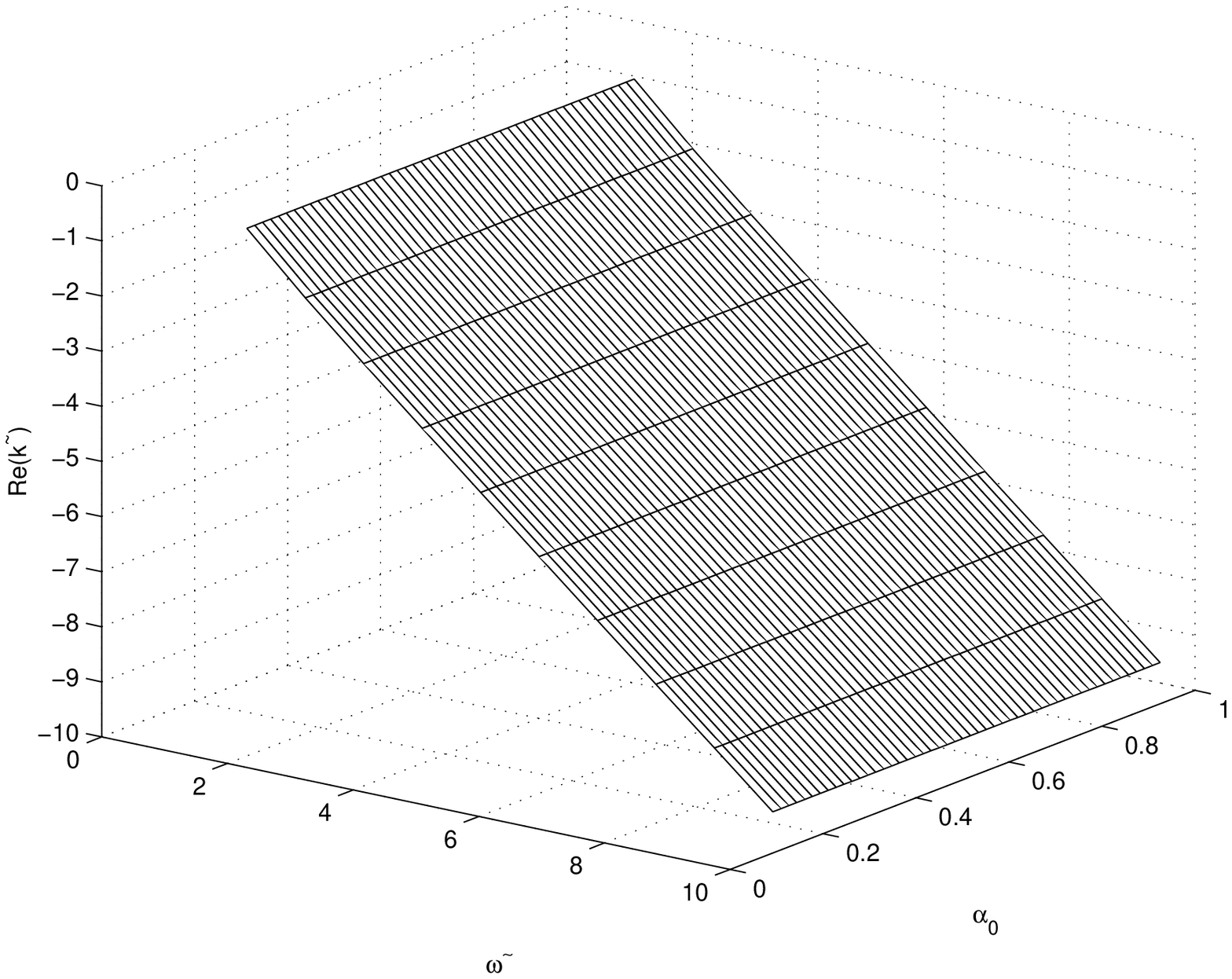}\\
\includegraphics[scale=0.38]{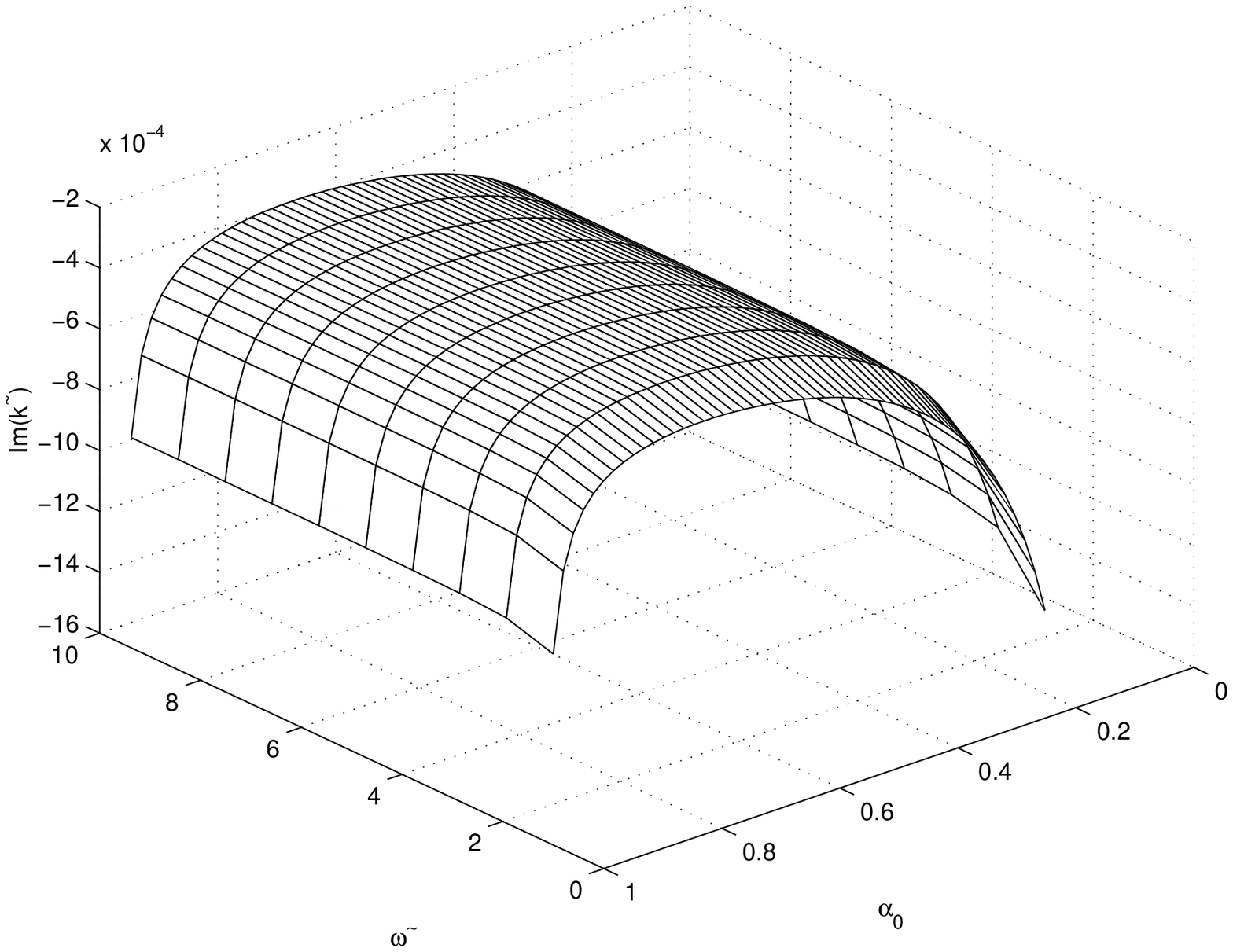}
\includegraphics[scale=0.38]{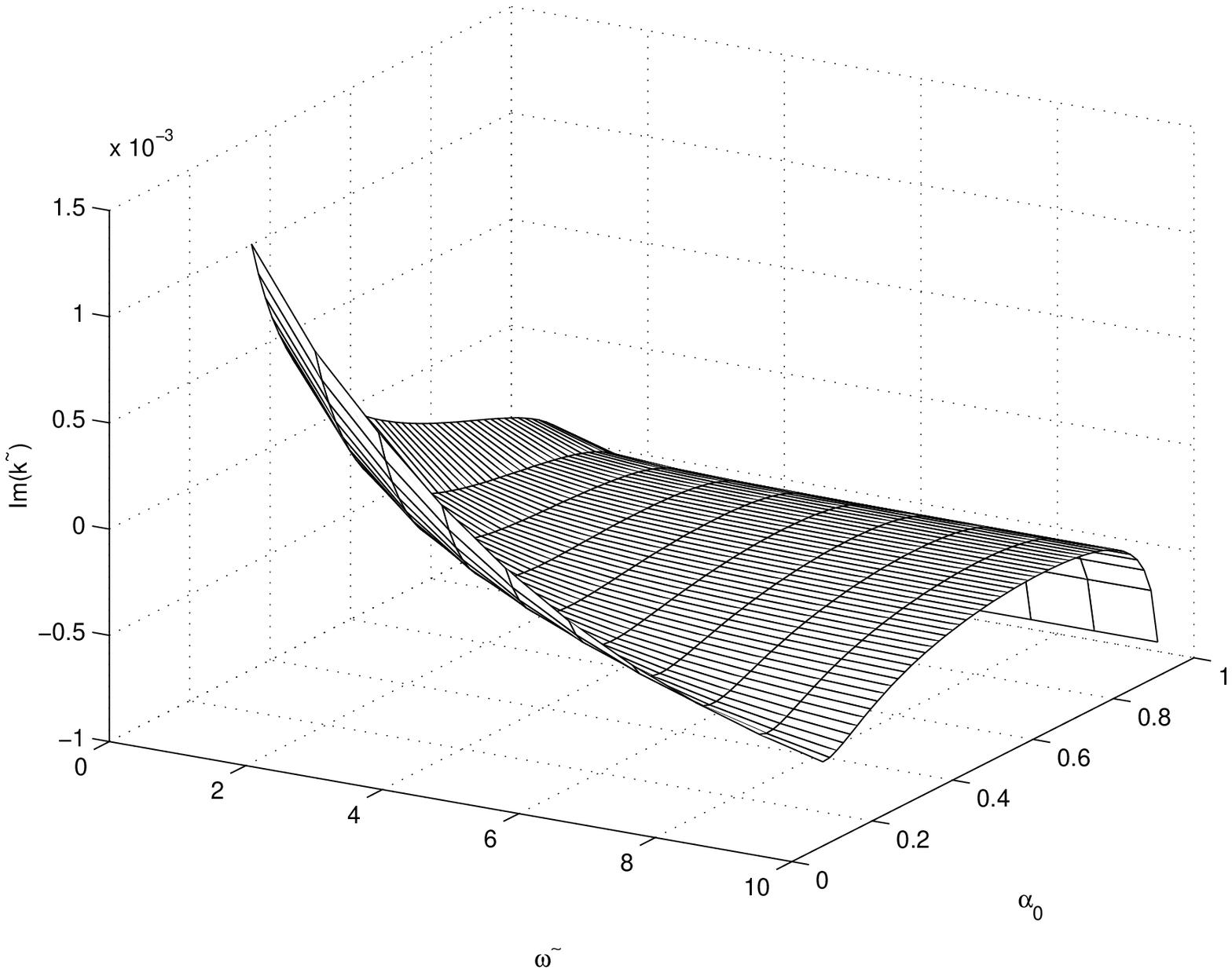}
\end{center}
\caption{\textbf{Top}: Real part of the growth mode and of the growth and damping mode for the high frequency transverse wave in the case of the electron-ion plasma. \textbf{Bottom}: \textit{Left}: Imaginary part of the growth mode. \textit{Right}: Imaginary part of the growth and damping mode.}
\end{figure}

\begin{figure}[h]\label{fig8}
\begin{center}
\includegraphics[scale=0.38]{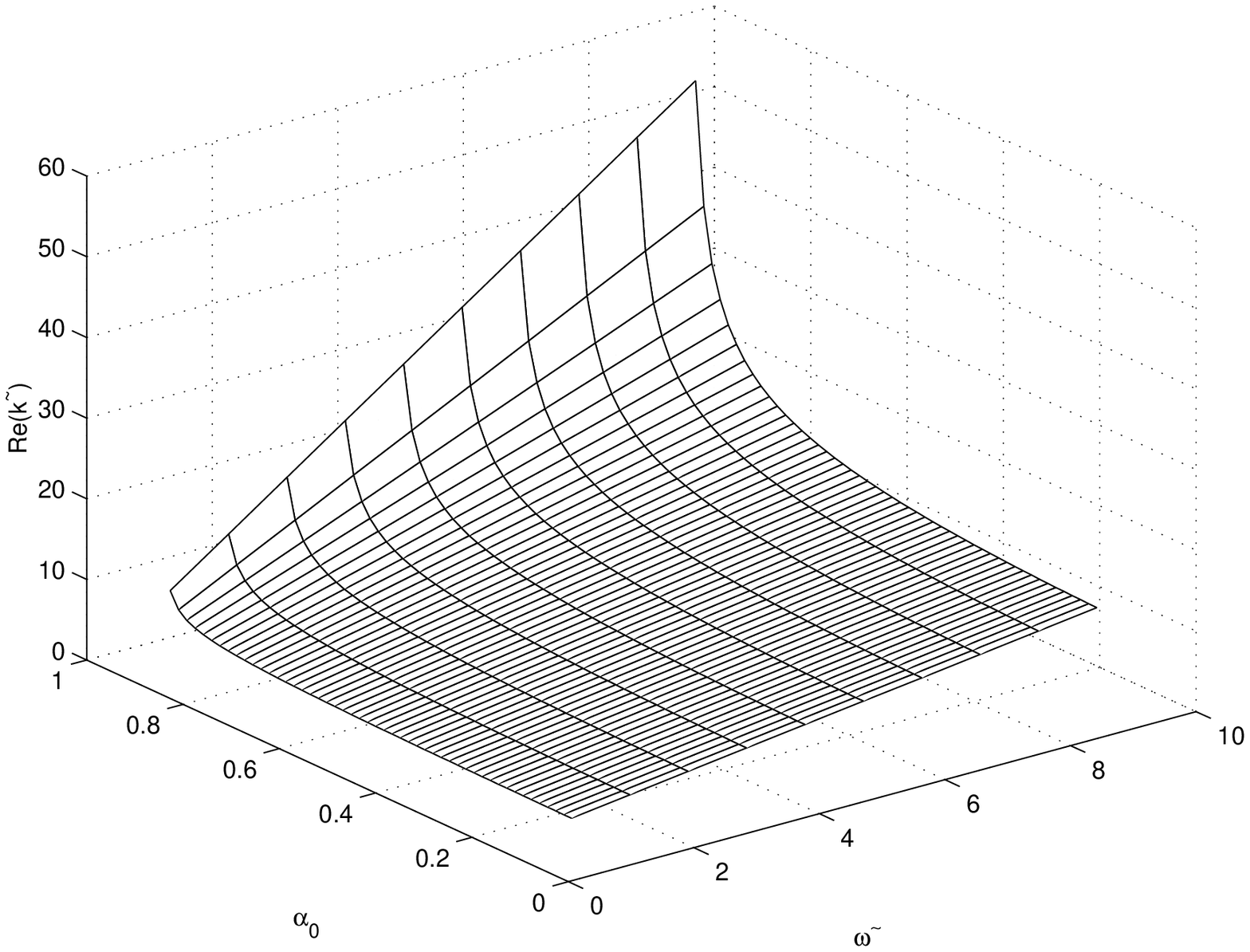}\\
\includegraphics[scale=0.38]{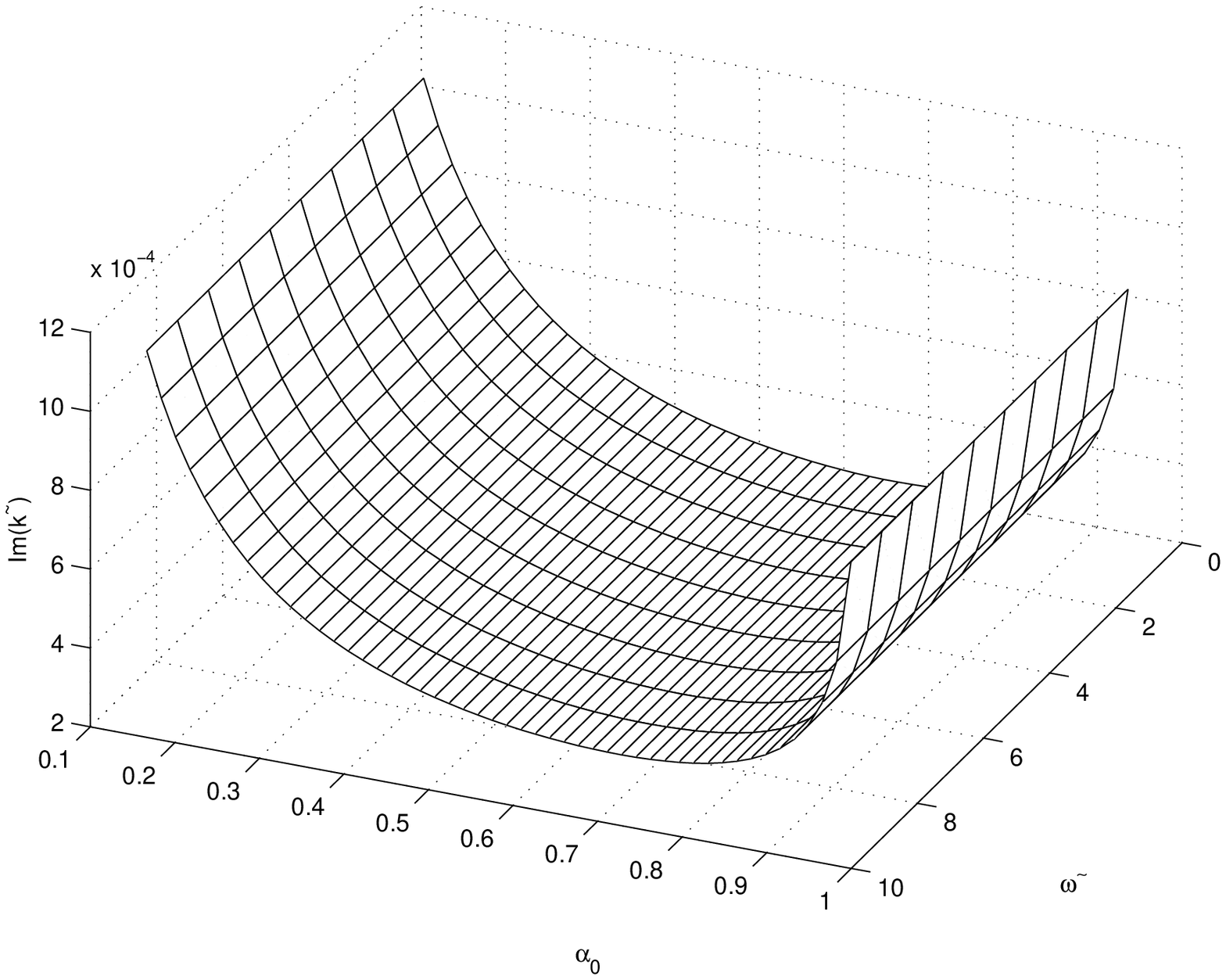}
\includegraphics[scale=0.38]{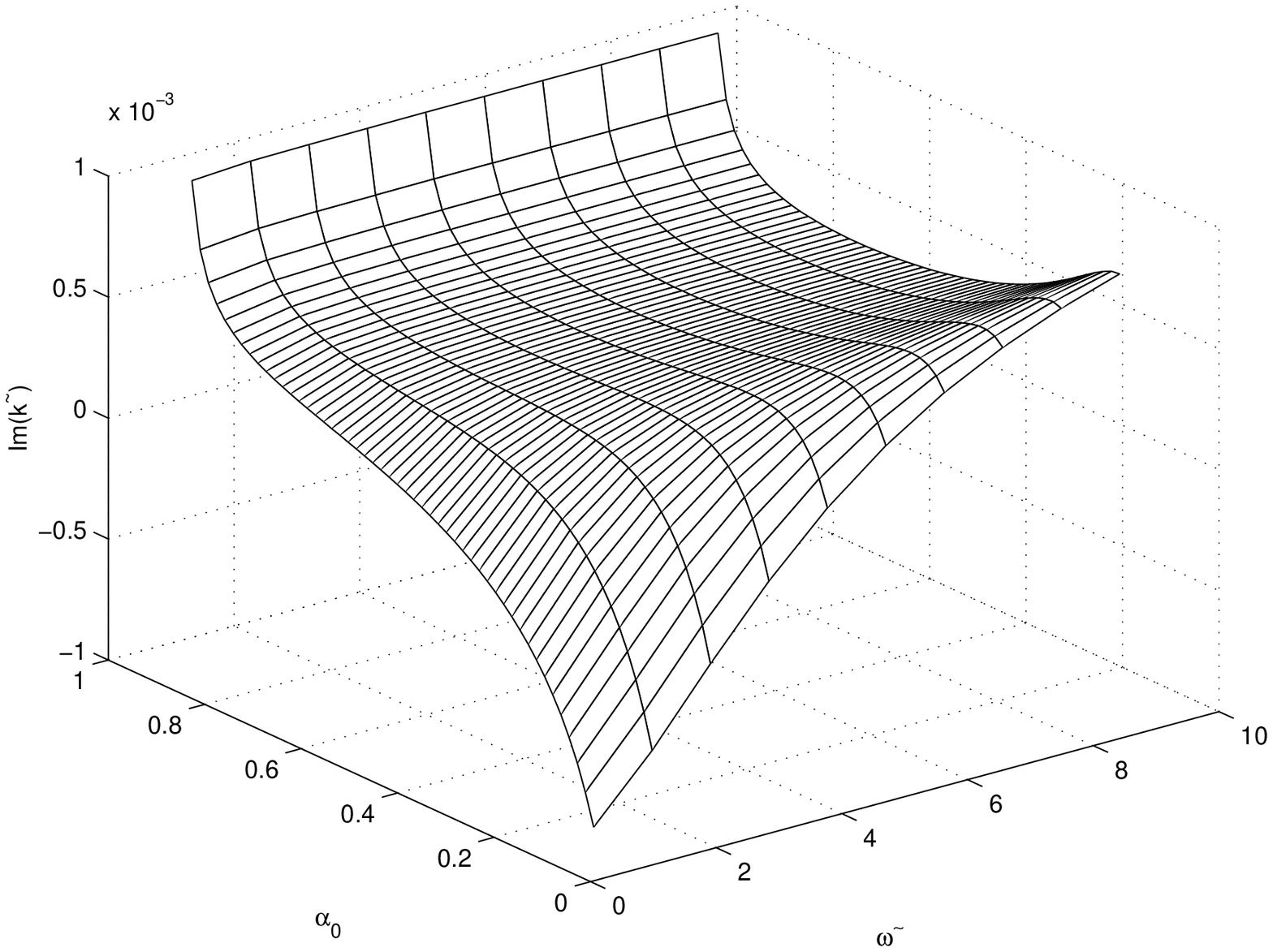}
\end{center}
\caption{\textbf{Top}: Real part of the damping mode and of the damping and growth mode for the high frequency transverse wave in the case of the electron-ion plasma. \textbf{Bottom}: \textit{Left}: Imaginary part of the one damped mode. \textit{Right}: Imaginary part of the other mode showing both damping and growth.}
\end{figure}

\subsubsection{\textit{Electron-ion Plasma}}\label{subsubsec8.1.2}

In the case of the electron-ion plasma there exist four modes, two of which, shown in Fig. 3, are complex conjugate pair and are significantly damped and growing respectively. The other two modes are also damping and growth. The third mode, shown in Fig. 4, is a growth mode. The fourth mode is similar to this mode but opposite in sign and is damped. The last two modes display only marginal damping and growth, respectively, and are equivalent to the two modes discussed in Fig. 2 for electron-positron plasma. For the first two modes, shown in Fig. 3, the differences in the magnitudes of the $\omega _{c1}$ and $\omega _{c2}$ apparently lead to take the frequencies from their negative (and therefore unphysical) values for the electron-positron case to positive physical values for the electron-ion case. These changes are because of the difference in mass and density factors as between the positrons and ions. These four modes coalesce with the modes of Buzzi et al. \cite{fourty seven} for the Schwarzschild case. It is evident that the growth and damping rates are independent of the frequency, but depended only on the radial distance from the black hole horizon through $\alpha _0$.

\subsection{High Frequency Transverse Modes}\label{subsec8.2}
\subsubsection{\textit{Electron-positron Plasma}}\label{subsubsec8.2.1}

There exist four high frequency electromagnetic modes for the electron-positron plasma. These are illustrated in Figs. 5 and 6. Fig. 5 shows two modes which are similar and both show a large amount of growth. The bottom left mode is clearly a growth mode for all the frequencies, but the bottom right mode is growth for higher frequency and is damped very close to the black hole horizon for lower frequency. Thus near the horizon, it appears that energy is no longer fed into wave mode by the gravitational field but begins to be drained from the waves.  The bottom left mode, shown in Fig. 6, is damped mode but the bottom right mode is growth for lower frequency as $\tilde \omega \rightarrow 1$ and $\alpha _0\rightarrow 0$. These four modes are similar to the three modes of Buzzi et al. \cite{fourty seven} with larger damping and growth rates.

\subsubsection{\textit{Electron-ion Plasma}}\label{subsubsec8.2.2}

Four high frequency modes are found to exist as shown in Figs. 7 and 8. The bottom left mode in Fig. 7 is a growth mode but the bottom right mode shows both growth and damping. For each of these modes, the solution become unstable for $\alpha _0\rightarrow 0$ and $\tilde\omega \rightarrow 1$. This may simply because the solution is too close to a resonant frequency. The damping and growth rates are larger than the corresponding modes for Schwarzschild case. Fig. 8 shows two modes, one of which shown in bottom left is a purely damped, while the bottom right one is both damping and growth. These modes are stable for all frequencies and at all distances from the horizon. Unlike the corresponding Alfv\'{e}n modes, the growth and decay rates obviously depend on frequency. The four modes coalesce with the three modes of Buzzi et al. \cite{fourty seven} for the Schwarzschild case.

\section{Concluding Remarks}\label{sec9}

The prime concern of this study has been exclusively the investigation, within the local approximation, of Alfv\'{e}n and high frequency transverse electromagnetic waves in a two-fluid plasma in the vicinity of the Reissner-Nordstr\"{o}m-de Sitter black hole.

We derive the dispersion relation for the Alfv\'{e}n and high frequency electromagnetic waves and solve it numerically for the wave number $k$. In the zero gravity limit, our study gives the results of the special relativistic case \cite{fifty three} where only one purely real mode was found to exist for both the Alfv\'{e}n and high frequency electromagnetic waves. In contrast to the special relativistic case, new modes (damped or growth) arise in the present work due to the black hole's gravitational field. We have found four Alfv\'{e}n and four high frequency modes either for electron-positron or electron-ion plasma. There is a variation of damping and growth rates of all the modes by some orders of magnitude with respect to the Schwarzschild case \cite{fourty seven}.

For the Alfv\'{e}n waves, the damping and growth rates are clearly independent of frequency for both the fluid components, but are solely dependent only on the radial distance from the horizon as denoted by the mean-value of the lapse function $\alpha_0$. On the other hand, for the high frequency electromagnetic waves the damping and growth rates are dependent on both frequency and radial distance from the black hole horizon. So, we may conclude that the damping modes demonstrate the energy drain from the waves by the gravitational field and the growth rates indicate that the gravitational field is, in fact, feeding energy into the waves.

The study of this paper gives the result for the lukewarm black hole when $q^2=M^2$. Moreover, the result can be specialized for the interesting cold, charged Nariai, and ultracold black holes by suitably choosing the black hole parameters as described in the introduction. In the limit $\ell^2\rightarrow\infty$ the study provides the result for the RN \cite{fifty seven} black hole, while for $\ell^2\rightarrow\infty$, $q^2=0$, the result reduces to that for the Schwarzschild black hole \cite{fourty seven}. The result for the Schwarzschild-de Sitter black hole \cite{fifty eight} is obtained if one sets $q^2=0$. Charged black holes in de Sitter space have interesting wormhole aspects \cite{fifty one,fifty nine}. In gravitational theory wormholes are bridges between different parts of the universe. There exists a class of extreme cases which appears to be classically and quantum mechanically stable \cite{fifty two,sixty}. The RNdS black hole is therefore interesting in a broader context. Our study of this paper is thus well motivated.

\vspace{1.0cm}

\noindent
{\large\bf Acknowledgement}\\
One of the authors (MHA) thanks the SIDA as well as the Abdus Salam International Centre for Theoretical Physics (ICTP), Trieste, Italy, for supporting with an Associate position of the Centre.

\end{document}